\documentclass[letterpaper,pdftex,rmp,twocolumn,amsmath,amssymb,groupedaddress,floatfix]{revtex4}
\usepackage{graphicx}
\RequirePackage[sort&compress]{natbib}
\usepackage{natbib}
\usepackage{color}
\usepackage{textcomp}
\usepackage{makecell}
\usepackage{mathptmx}
\usepackage{float}
\usepackage{setspace}
\usepackage[a]{esvect}
\usepackage[font=sf,size=small,justification=RaggedRight,singlelinecheck=false]{caption}
\DeclareCaptionLabelSeparator{bar}{ \rule[-0.2em]{0.3ex}{1em} }
\usepackage[sf,bf,small]{titlesec}
\titlespacing*{\section}{0pt}{1em}{0em}

\definecolor{darkgray}{rgb}{0.25,0.25,0.25}
\definecolor{darkred}{rgb}{0.89,0.10,0.11}
\definecolor{darkblue}{rgb}{0.12,0.39,0.62}
\usepackage{url}

\newcommand{\hs}{\@~}

\usepackage[pdftex,breaklinks=true,colorlinks=true,citecolor=black,linkcolor=black,menucolor=black,urlcolor=darkblue,pdfborder={1 0 0}]{hyperref}
\hypersetup{pdftitle={Memory in network flows and its effects on spreading dynamics and community detection},pdfauthor={M. Rosvall, A. Viamontes Esquivel, A. Lancichinetti, J.D. West, and R. Lambiotte (2014)}}
\bibpunct{}{}{,}{s}{,}{,}

\usepackage{booktabs}

\newcommand{\mytoprule}{\specialrule{0.1em}{0em}{0em}}
\newcommand{\mybottomrule}{\specialrule{0.1em}{0em}{0em}}

\begin{document}
\makeatletter
\renewcommand\@biblabel[1]{#1.}
\makeatother
	
\renewcommand{\figurename}{Figure}
\renewcommand{\thefigure}{\arabic{figure}}
\renewcommand{\tablename}{Table}
\renewcommand{\thetable}{\arabic{table}}
\renewcommand{\refname}{\large References}

\addtolength{\textheight}{1cm}
\addtolength{\textwidth}{1cm}
\addtolength{\hoffset}{-0.5cm}

\setlength{\belowcaptionskip}{1ex}
\setlength{\textfloatsep}{2ex}
\setlength{\dbltextfloatsep}{2ex}

\hyphenation{page-rank}

\newcommand*{\citen}[1]{%
  \begingroup
    \romannumeral-`\x 
    \setcitestyle{numbers}%
    \cite{#1}%
  \endgroup   
}

\title{Memory in network flows and its effects on spreading dynamics and community detection}


\author{Martin Rosvall}
\affiliation{Integrated Science Lab, Department of Physics, Ume{\aa} University, SE-901 87 Ume{\aa}, Sweden}

\author{Alcides V. Esquivel}
\affiliation{Integrated Science Lab, Department of Physics, Ume{\aa} University, SE-901 87 Ume{\aa}, Sweden}

\author{Andrea Lancichinetti}
\affiliation{
Howard Hughes Medical Institute (HHMI) \& 
Department of Chemical and Biological Engineering,
Northwestern University, Evanston, Illinois 60208 USA}
\affiliation{Integrated Science Lab, Department of Physics, Ume{\aa} University, SE-901 87 Ume{\aa}, Sweden}

\author{Jevin D. West}
\affiliation{Integrated Science Lab, Department of Physics, Ume{\aa} University, SE-901 87 Ume{\aa}, Sweden}
\affiliation{Information School, University of Washington, Seattle, Washington 98195, USA}

\author{Renaud Lambiotte}
\affiliation{Department of Mathematics and Naxys, University of Namur, 5000 Namur, Belgium}

\begin{abstract}
Random walks on networks is the standard tool for modelling spreading processes in social and biological systems. This first-order Markov approach is used in conventional community detection, ranking, and spreading analysis although it ignores a potentially important feature of the dynamics: where flow moves to may depend on where it comes from. Here we analyse pathways from different systems, and while we only observe marginal consequences for disease spreading, we show that ignoring the effects of second-order Markov dynamics has important consequences for community detection, ranking, and information spreading.
For example, capturing dynamics with a second-order Markov model allows us to reveal actual travel patterns in air traffic and to uncover multidisciplinary journals in scientific communication. These findings were achieved only by using more available data and making no additional assumptions, and therefore suggest that accounting for higher-order memory in network flows can help us better understand how real systems are organized and function.
\end{abstract}

\maketitle

\noindent A central objective of network science is to connect structure with dynamics in integrated social and biological systems \cite{watts1998collective,barabasi1999emergence,barrat2004architecture,boccaletti2006complex}.
In this data-driven approach, the complex structure is represented with a network of nodes and links, and the dynamics are modelled with random flow on the network \cite{granovetter1983strength,brin1998anatomy,brockmann2006scaling,balcan2009multiscale,vespignani2012modelling}.
The flow can represent ideas circulating among colleagues, passengers travelling through airports, or patients moving between hospital wards. Conventional network models implicitly assume that where the flow moves to only depends on where it is, and that this first-order Markov process suffices for performing community detection, ranking, and spreading analysis. Claude Shannon introduced higher-order memory models in 1948 \cite{shannon1948mathematical}, and there is a substantial body of work on analysing memory effects in, for example, time-series analysis for forecasting financial markets \cite{box2013time}, correlated random walks for predicting animal movements \cite{kareiva1983analyzing}, and exponential random graph models for capturing social networks \cite{robins2007recent}. Moreover, there is recent evidence that memory is necessary for accurately predicting web traffic \cite{meiss2008ranking,chierichetti2012web}, for improving search and navigation in information networks \cite{asztalos2010network,backstrom2011supervised,singer2014memory},
 and for capturing important phenomena in the spread of information \cite{iribarren2009impact,takaguchi2011predictability,holme2012temporal,lentz2013unfolding,pfitzner2013betweenness} and epidemics \cite{gonzalez2008understanding,heath2008construction,song2010limits,balcan2011phase,belik2011natural,poletto2013human}. Nevertheless, little is known about memory effects on community detection, ranking, and spreading analysis, three principal methods in network science. This issue raises a fundamental question that allows us to better understand social and biological systems: what are the effects of ignoring higher-order memory in network flows on community detection, ranking, and spreading?

\begin{figure}[ht]
  \centering
  \includegraphics[width=\columnwidth]{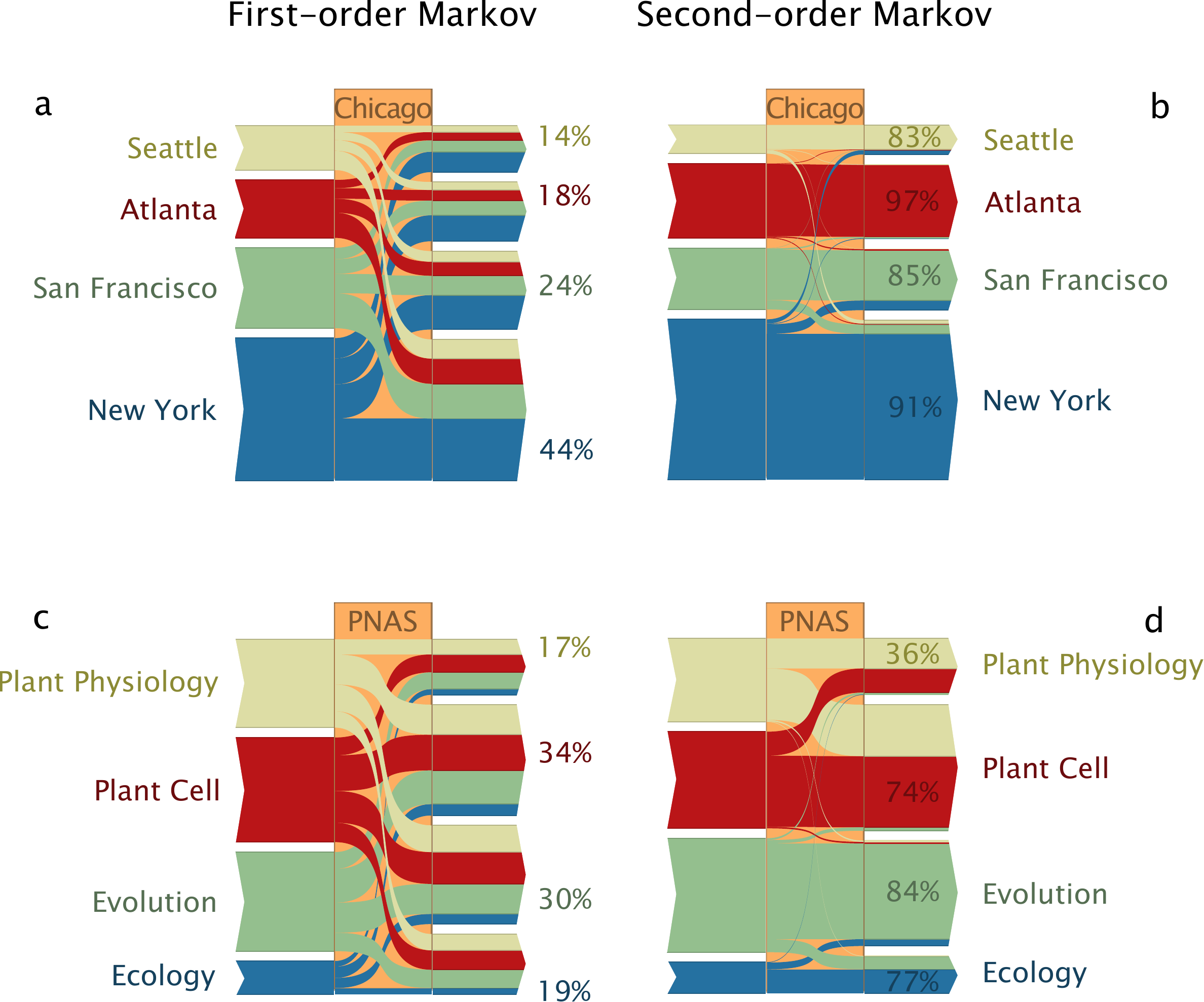}
  \caption{\textbf{First-order Markov dynamics distort real constraints on flow.}
  (\textbf{a}) In a first-order Markov approach, we model passengers' travel to a city to be proportional to the observed volume of traffic to that city, \emph{and irrespective of where the passengers come from}.
  (\textbf{b}) In a second-order Markov model, passengers' travel to a city is still proportional to the traffic volume, \emph{but also dependent on where the passengers come from}.
  In this example, out-and-back traffic to Chicago only dominates over transfer traffic when second-order Markov dynamics are taken into account.
  (\textbf{c}-\textbf{d}) Journal citation flow shows the same memory effect. Citation flow from four different journals to \emph{PNAS} is mostly shown to return to the same journal or continue to a related journal only when second-order Markov dynamics are taken into account.
  The percentages represent the relative return flow.\label{socialflow}}
\end{figure}

To comprehend the effects of memory, we use networks in which the direction of flow depends on the weights of the outgoing links and, importantly, \emph{where the flow comes from}. In this paper, we focus on second-order Markov dynamics such that the next step depends on the currently and previously visited node, which corresponds to a second-order Markov model of flow.
As an illustration, we use air traffic between airports of different cities with link weights derived from real itineraries (Fig.\ \ref{socialflow}).
When we take first-order Markov dynamics into account in the conventional network approach, nodes $i$ represent cities and links $i \rightarrow j$ represent flight legs, with weights $W(i \rightarrow j)$ proportional to the passenger volume between cities.
The dynamics are modelled with weighted steps between nodes on \emph{networks without memory} and correspond to a first-order Markov model of flow, since the direction of flow only depends on the currently visited city (Fig.\ \ref{socialflow}a). This conventional approach is used in a wide range of problems, from ranking nodes \cite{brin1998anatomy} and finding communities \cite{rosvall2008maps,delvenne2010stability} to modelling the spread of epidemics \cite{may2001infection,pastor2001epidemic} and rumours \cite{nekovee2007theory}.
However, this approach ignores where the passengers come from, and therefore the direction of passenger flow is independent of the incoming traffic.
When we take second-order Markov dynamics into account, on the other hand, \emph{memory nodes} $\vv{ij}$ represent flight legs and links $\vv{ij} \rightarrow \vv{jk}$ represent connected flight legs, with weights $W(\vv{ij} \rightarrow \vv{jk})$ proportional to the passenger volume between cities and conditional on the previously visited city. In this way, a city is represented by a \emph{physical node} $j$ with multiple \emph{memory nodes} $\vv{ij}$, one for each incoming flight leg from city $i$, such that arriving in Chicago from Seattle corresponds to arriving at memory node $\vv{Seattle\;Chicago}$ of physical node $Chicago$. By modelling the dynamics on this \emph{network with memory}, such that steps depend on the currently and previously visited city, we can better reveal actual travel patterns (Fig.\ \ref{socialflow}b).

While we considered passengers moving between cities in this illustration, we have analysed six diverse systems in detail, including researchers navigating between journals and patients moving between wards. We find that taking second-order Markov dynamics into account is important for understanding the actual dynamics, because random dynamics on networks obscure essential structural information. After deriving how we model the dynamics and quantify their constraints, we show how the second-order Markov constraints on dynamics influence three important branches of network science: community detection, ranking, and epidemic spreading.

\section*{Results}
\noindent\textbf{Modelling second-order Markov dynamics.} 
For each system, we model the dynamics as a stochastic process. We represent the $n$ different components of the system with physical nodes $i=1,2,\ldots,n$ and let $X_t$ denote the state or position of an entity of flow at time $t$. With this notation, the flow through the system corresponds to a walker stepping between nodes, which can be described by an indexed sequence of random variables $X_{1} X_{2} \ldots X_{t}$.
In general, the probability that the flow visits node $i$ at time $t+1$ depends on the full history of the dynamic process:
\begin{align}
P(i;t+1) &\equiv P(X_{t+1} = i_{t+1})\\
		&= P(X_{t+1} = i_{t+1} | X_{t} = i_{t}, X_{t-1} = i_{t-1}, \ldots, X_{1} = i_{1} ), \nonumber
\end{align}
for all $i_1,i_2,\ldots, i_t,i_{t+1} \in i=1,2,\ldots,n$.
In network science, it is common to assume that the direction flow takes in a dynamic process depends only on the current state and not on time: 
\begin{align}
P(i;t+1) = P(X_{t+1} = i_{t+1} | X_{t} = i_{t}) = P(X_{2} = i_{t+1} | X_{1} = i_{t}).
\end{align}
In other words, the dynamic process is Markovian or a first-order Markov process (M1). That is, it is assumed that knowledge about the relative weights of links between the nodes is sufficient to model the dynamic process in the system. All this information is captured in the transition matrix $P$ with elements of the form
\begin{align}
P_{ij} = p(i \to j) = \frac{W(i \to j)}{\sum_k W(i \to k)},\label{firstordertransitionprob}
\end{align}
measuring the probability that a random walker at node $i$ steps to node $j$ and normalized such that $\sum_j p(i \to j) = 1$. Accordingly, the probability of finding the random walker at node $j$ at time $t+1$ is 
\begin{align}
P(j;t+1) = \sum_i P(i;t) p(i \to j).
\end{align}
Many ranking \cite{brin1998anatomy,bergstrom2008eigenfactor} and community detection \cite{rosvall2008maps,delvenne2010stability} methods as well as epidemic models \cite{vespignani2012modelling} build directly on this first-order Markov process. In fact, also maximal-entropy random walks are Markovian, though they build on modified transition probabilities \cite{parry1964intrinsic,sinatra2011maximal}.

As we argue below, random dynamics on networks cannot accurately capture empirical flow pathways. As a result, a first-order Markov modelling can fail to capture important phenomena in a broad range of complex systems \cite{may2001infection,pastor2001epidemic}. To capture higher-order Markov effects in flow pathways \cite{gonzalez2008understanding,song2010limits,balcan2011phase,belik2011natural}, we use memory networks. A memory network consists of memory nodes; each memory node represents the current state of the walker, the currently visited node, and the previous step or steps. The order of the Markov process determines the number of steps that represent a state. For example, in a second-order Markov process (M2), the walker's next step depends on the currently visited node $j$ and the previously visited node $i$. In this case, the memory nodes $\vv{ij}$ correspond to directed links between physical nodes in the standard network. Accordingly, the network of memory nodes is a form of line graph of the network without memory (see the Supplementary Information, Sec.\ 1). In a third-order Markov process, the walker's next step depends on the currently visited node $j$ and the two previously visited nodes $h$ and $i$, and the memory nodes $\vv{hij}$ correspond to three-step pathways between physical nodes in the standard network. Here we focus on a second-order Markov process, but the procedure can in principle easily be generalized to higher-order Markov processes, provided that sufficient data is available to fit the model. 

The dynamics of a second-order Markov walker can now simply be modelled as a Markov process on the memory network, instead of a non-Markov process on the physical nodes. For a second-order Markov process, the dynamics are encoded by a transition matrix with elements of the form
\begin{align}
\label{aa}
p(\vv{ij} \rightarrow \vv{jk}) = \frac{W(\vv{ij} \to \vv{jk})}{\sum_{l} W(\vv{ij} \to \vv{jl})},
\end{align}
measuring the probability that the walker steps from $j$ to $k$ if it came from $i$ in the previous step and normalized such that $\sum_k p(\vv{ij} \rightarrow \vv{jk}) = 1$.  These transitions can therefore be interpreted as movements between links.  However, even in undirected networks, we must use two memory nodes for each pair of connected nodes $i$ and $j$ since the memory nodes encode the time ordering of the visits.
In any case, the probability of finding the random walker at memory node $\vv{jk}$ at time $t+1$ is
\begin{align}
\label{general1}
P(\vv{jk};t+1) = \sum_{i} P(\vv{ij};t) p(\vv{ij} \rightarrow \vv{jk}).
\end{align}
Consequently, the probability of finding the random walker at physical node $k$ at time $t+1$ in a second-order Markov process is
\begin{align}
P(k;t+1) = \sum_j P(\vv{jk};t+1) = \sum_{ij} P(\vv{ij};t) p(\vv{ij} \rightarrow \vv{jk}).
\end{align}

\noindent\textbf{Constraints on flow captured in real-world pathway data.} 
We collected pathway data with sequences of steps for the six well-studied and diverse systems presented in Table~\ref{table:stats}: flight itineraries between US \emph{airports}, the airports aggregated in \emph{cities}, chains of citing articles aggregated in \emph{journals}, movements of \emph{patients} between hospital wards in Stockholm, GPS-tracked \emph{taxis} in San Francisco, and chains of forwarded and replied \emph{emails} (see the Supplementary Information, Sec.\ 1). We chose these systems because their pathway data were readily available and because the outcomes of their analyses have important consequences. To explain the effects of memory, we analysed the systems with networks with and without memory.

\begin{table*}[ht]
\caption{\textbf{Summary of second-order Markov effects in real-world networks}\label{table:stats}}\vspace{-1em}
\flushleft
\setlength{\tabcolsep}{2.9pt}
\setstretch{0.85}
\begin{small}
\begin{tabular*}{0.88\hsize}{@{}rrrrrllllllllllllllllll@{}}\mytoprule\noalign{\smallskip}
{Network} &\rule{0.5em}{0em}& \multicolumn{2}{l}{ Number of} &\rule{0.5em}{0em}& \multicolumn{2}{l}{ Two-step } &\rule{0.5em}{0em}& \multicolumn{2}{l}{ Three-step  } &\rule{0.5em}{0em}& \multicolumn{2}{l}{ Entropy } &\rule{0.5em}{0em}& \multicolumn{2}{l}{ Module } &\rule{0.5em}{0em}& \multicolumn{2}{l}{ Module } &\rule{0.5em}{0em}& \multicolumn{1}{l}{Compression} &\rule{0.5em}{0em}& Ranking \\
{ } && \multicolumn{2}{l}{nodes} && \multicolumn{2}{l}{ return (\%)} && \multicolumn{2}{l}{ return (\%)} && \multicolumn{2}{l}{rate (bits)} && \multicolumn{2}{l}{size (\%)} && \multicolumn{2}{l}{ assignmt.} && gain~(\%) && diff.~(\%) \\ \noalign{\smallskip}
&& { M1} & { M2} && { M1} & { M2} && { M1} & { M2} && { M1} & { M2} && { M1} & { M2} && { M1} & { M2} && { M1\raisebox{0.1ex}{$\to$}M2} && { M1\raisebox{0.1ex}{$\to$}M2} \\ \noalign{\smallskip}
\cline{0-0}\cline{3-4}\cline{6-7}\cline{9-10}\cline{12-13}\cline{15-16}\cline{18-19}\cline{21-21}\cline{23-23}
\noalign{\smallskip}
Airports && 464 & 17,983 && 5.7 & 47 && 2.1 & 0.63 && 5.2 & 3.4 && 93 & 5.1 && 1.2 & 6.2 && \rule{1.5em}{0em}13 && \rule{1.4em}{0em}8.2 \\
Cities && 413 & 15,368 && 6.5 & 48 && 2.8 & 0.62 && 4.7 & 3.5 && 32 & 5.3 && 1.8 & 3.7 && \rule{1.5em}{0em}5.2 && \rule{1.5em}{0em}3.7\\
Journals && \rule{-1em}{0em}1,983 & 201,349 && 11 & 21 && 4.7 & 5.4 && 4.5 & 3.5 && 14 & 15 && 1.8 & 3.4 && \rule{1.5em}{0em}4.7 && \rule{1.5em}{0em}9.7\\
\noalign{\smallskip}
Patients && 402 & 4,987 && 16 & 54 && 1.9 & 3.4 && 3.0 & 1.0 && 7.3 & 1.9 && 5.0 & 4.7 && \rule{1.5em}{0em}30 && \rule{1.5em}{0em}22\\
Taxis && 416 & 2,763 && 20 & 10 && 6.8 & 10 && 2.2 & 1.1 && 3.1 & 2.2 && 1.5 & 1.7 && \rule{1.5em}{0em}7.1 && \rule{1.5em}{0em}6.5\\
Emails && 144 & 1,432 && 14 & 58 && 5.2 & 2.7 && 3.0 & 1.3 && 12 & 5.8 && 1.3 & 3.0 && \rule{1.5em}{0em}26 && \rule{1.5em}{0em}18\\
\mybottomrule\noalign{\smallskip}
\end{tabular*}
\parbox{\textwidth}{\footnotesize\raggedright In Supplementary Tables 1 and 2, we provide 10th and 90th percentiles from bootstrap analysis.}
\end{small}
\end{table*}

\noindent\textbf{Memory dynamics better reveal real constraints on flow.}
The pathways in Fig.\ \ref{socialflow} illustrate how second-order Markov dynamics strongly direct flow in two real-world examples. With data from actual itineraries, Figs.\ \ref{socialflow}a-b show trips to/from Chicago modelled with first-order Markov dynamics in a and with second-order Markov dynamics in b (see Methods). When only the relative proportions of departures from Chicago determines the next destination in the standard network representation, the trips mix randomly. With a second-order Markov model, however, passengers flying to Chicago are most likely to return to the city from which they came.
Similarly, Figs.\ \ref{socialflow}c-d show the journal citation flow to/from the journal \emph{PNAS} with first-order Markov dynamics in c and with second-order Markov dynamics in d. The journal citation flow is a proxy for how researchers navigate scholarly literature, derived from a random walker moving between articles following citations and mapped onto journals. 
When only the fraction of citations from \emph{PNAS} to the specialized journals determines which journal the walker reads next, the pathways mix randomly. Instead, with second-order Markov dynamics taken into account, after following a citation in an article published in a more specialized journal to an article in \emph{PNAS}, the walker tends to return to an article published in the same specialized journal or field. Defined as the relative amount of flow that returns to the same physical node after two steps, the two-step return rate is twice as large when second-order Markov dynamics is accounted for in citation flow and eight times as large in passenger flow. Except for the taxi data (taxis take us to destinations away from where we were), we found that second-order Markov dynamics reveal a dramatically higher return flow in all studied systems (Table~\ref{table:stats}).

\begin{figure}[ht]
    \centering
   \includegraphics[width=\columnwidth]{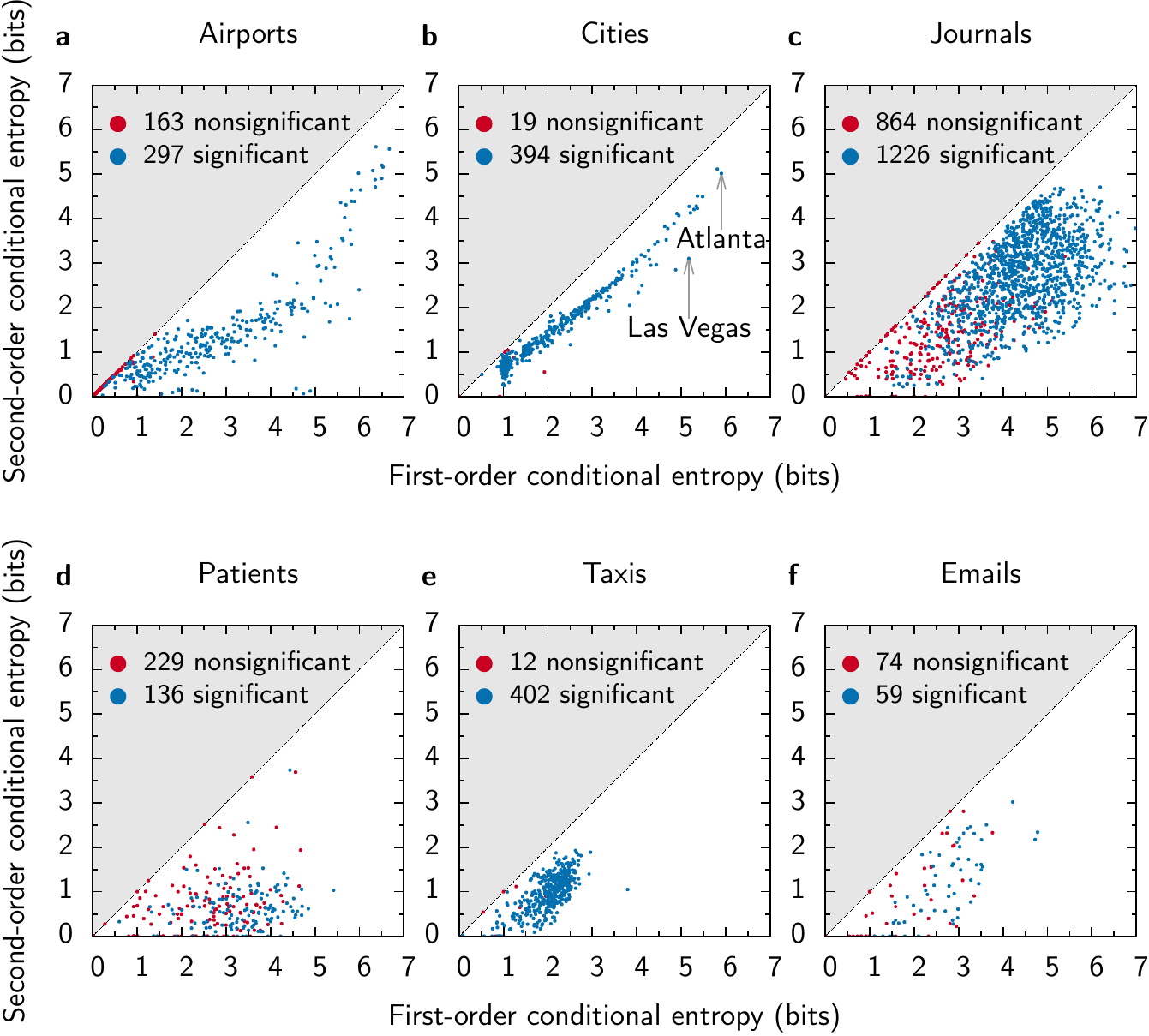}
    \caption{\textbf{Significant second-order Markov constraints on flow.} (\textbf{a}-\textbf{f}) First- and second-order conditional entropy for all nodes of the six analyzed networks. Blue nodes show a significant memory effect, because the null hypothesis that the data are generated from a first-order Markov model can be rejected. Red nodes do not show a significant effect. The memory effect is the difference in entropy between a first- and second-order Markov model. Las Vegas, among all cities, shows the strongest memory effect. Traffic is dominated by visitors who return to the city from which they came. In the other extreme, nodes that we could not significantly distinguish from a first-order Markov model typically have low connectivity and relatively small entropies.\label{markovtest}}
\end{figure}

To quantify the second-order Markov constraints on flow, we measured the entropy rate of a random walker on a network with and without memory \cite{shannon1948mathematical}.
The entropy rate measures the conditional entropy, the uncertainty of the next step of the flow given the current state, weighted by the stationary distribution. In a first-order Markov process, the entropy rate is the conditional entropy at each physical node weighted by the stationary distribution:
\begin{align}
H(X_{t+1}|X_{t}) = -\sum_{jk}\pi(j) p(j \to k) \log{ p(j \to k) },
\end{align}
where $\pi$ is the stationary solution of the random process. In a second-order Markov process, the entropy rate is the conditional entropy at each memory node weighted by the stationary distribution:
\begin{align}
H(X_{t+1}|X_{t}X_{t-1}) = -\sum_{ijk}\pi(\vv{ij}) p(\vv{ij} \to \vv{jk}) \log p(\vv{ij} \to \vv{jk}).
\end{align}
The more effect memory has, the more the conditional entropy will decrease in the second-order Markov model.
For the analysed networks, the entropy rates decrease by one to two bits when second-order Markov dynamics are taken into account (see Table~\ref{table:stats}). To put this decrease in perspective, we can compare with an unweighted network, in which the typical number of neighbours halves for each bit the entropy rate decreases.
That is, were the links unweighted, the observed decrease in entropy rates would correspond to overestimating the effective number of neighbours by 200\%--400\%. The nodes with the strongest memory effect have high entropy with first-order and low entropy with second-order Markov dynamics. For many nodes, memory greatly reduces the effective connectivity and reveals the constraints on flow (Fig.\ \ref{markovtest}).

\begin{figure}[ht]
    \centering
   \includegraphics[width=\columnwidth]{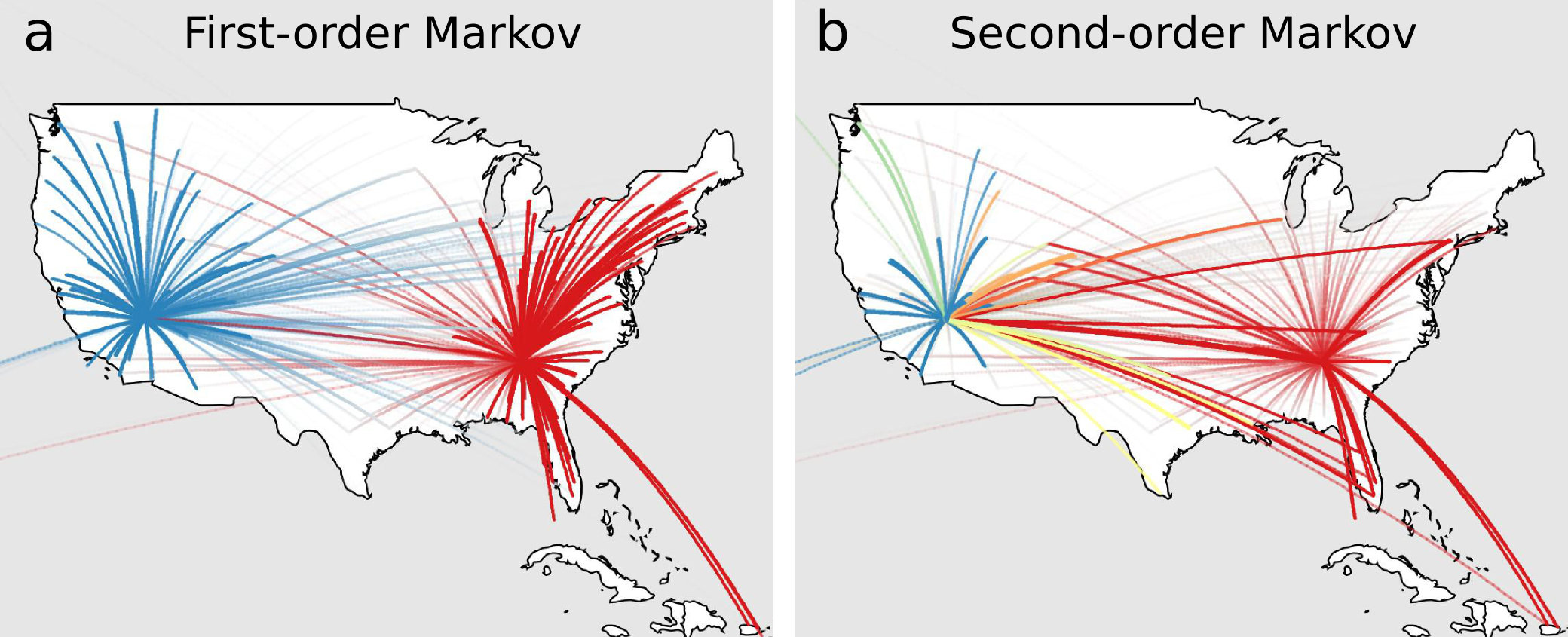}
    \caption{\textbf{Memory affects modular overlap in air traffic between US cities.} Major modules of Las Vegas and Atlanta with first-order Markov dynamics in \textbf{a} and second-order Markov dynamics in \textbf{b}. Link colours represent modules and link thicknesses represent passenger volume. \label{memoryairclustering}}
\end{figure}

Second-order Markov constraints on flow are statistically significant. To verify that our results are based on sufficient data, we performed bootstrap resampling of pathways for all summary statistics and surrogate data testing of the entropy rate to estimate the Markov order \cite{van1998testing} (see Fig.\ \ref{markovtest}, Methods, and the Supplementary Information, Sec.\ 2).  All summary statistics in Table~\ref{table:stats} and a majority of influential nodes in all networks except patients and emails show a significant second-order Markov effect that cannot be explained by noise.
While we focus on second-order Markov dynamics in this paper, it is interesting to reflect on potential effects of higher-order Markov models. For example, a second-order Markov model captures real dynamics with one-step memory, including the two-step return rate, a third-order Markov model captures two-step memory, including the three-step return rate, and so on. In principle, we could go to any order $n$ for higher accuracy. In practice, however, higher-order Markov models are more complex and demand many long pathways to statistically separate real effects of memory from insufficient data \cite{chierichetti2012web}. For the air-traffic data, we have enough long pathways to measure the entropy rate of a higher-order Markov model. When we estimated the average amount of information necessary to determine the next destination of passengers at airports, we measured a 0.3 bit decrease from second to third order compared with 1.1 bits from first to second order (see the Supplementary Information, Sec.\ 2 and Supplementary Fig.\ 3). While both results are statistically significant according to a surrogate data test, this small difference suggests that a second-order Markov model captures most of the salient features set by the constraints on flow in air-traffic, namely, that passengers tend to return to the city from which they came. 

We now turn to the consequences of ignoring higher-order memory when analysing network flows in social and biological systems. To study the consequences, we modified and generalized three commonly used network techniques to capture the effects of memory in a second-order Markov model: the map equation for community detection, PageRank for ranking, and two compartmental models for spreading. We begin with community detection, since simplifying and highlighting important structures of the dynamics allow us to better understand and explain the effects of second-order Markov dynamics on ranking and spreading dynamics.\\

\noindent\textbf{Memory affects community detection.} We used the map equation framework to identify overlapping modules with long flow persistence times \cite{rosvall2008maps,esquivel2011compression} in networks with and without memory (see Methods and the Supplementary Information, Sec.\ 3). This information-theoretic method measures how efficient a modular description is in compressing the pathways of a random walker. The more structural information that can be exploited, the better the compression \cite{shannon1948mathematical}. 
We measured how well modules identified with first- and second-order Markov dynamics can compress the more detailed model of the actual pathways (see Methods).
Table~\ref{table:stats} shows that second-order Markov dynamics allow for better compression, because random dynamics on networks obscure essential structural information. We quantified this structural information in terms of module size and level of module overlap. Measured as the average visit frequency of a random walker in each module, and weighted by the same visit frequency to reduce the effect of small modules, we report the effective module size for all networks in Table~\ref{table:stats}. Community detection of passenger traffic modelled as first-order Markov dynamics only identifies major geographic regions, such as the West, the South, the Mid-West, and the East. Second-order Markov dynamics reveal much more detailed travel patterns and the typical module size is more than five times smaller.
Analysing the hospital data, we found that patients are sent back to the previously visited ward more than half of the time, or more than three times as often as asserted by a standard network approach. As a result, the typical module within which patients move is significantly smaller when second-order Markov effects are taken into account. Memory also impacts information spreading through email communication. We found that the two-step return rate was four times higher with second-order Markov dynamics, thus revealing an organization with halved module sizes. We used the map equation framework because it was straightforward to generalize its mathematics to second-order Markov dynamics, but the results are, in principle, universal for any method operating on the dynamics on a network \cite{delvenne2010stability}. The universality is manifested in the direct effect memory in network flows has on the spectral gap \cite{scholtes2013slow,lambiotte2014effect}. If memory favours spread across a system, the spectral gap increases, and, the other way around, if memory confines flow, the spectral gap decreases. Overall, in the systems analysed here, second-order Markov dynamics reveal a higher return flow that confines flow in smaller and more informative modules.


Memory affects the level of module overlap. In air traffic between US cities modelled with first-order Markov dynamics, both Las Vegas and Atlanta are assigned to a single major module, as shown in Fig.\ \ref{memoryairclustering}a, but second-order Markov dynamics reveal their different flow patterns. Atlanta, with many transferring passengers and a relatively low two-step return rate (15\% with second-order and 1.8\% with first-order Markov dynamics),
is assigned to only one major module shown in red in Fig.\ \ref{memoryairclustering}b. In contrast, Las Vegas, with traffic dominated by returning tourists (67\% two-step return rate with second-order and 3.7\% with first-order Markov dynamics), is assigned to eight major modules, as shown in Fig.\ \ref{memoryairclustering}b (see Methods). Similarly, Supplementary Table~3 shows that second-order Markov dynamics can reveal multidisciplinary journals in the scholarly literature. For example, an ecologist reading an article published in \emph{PNAS} will most likely next read an article published in an ecological journal, as shown in Fig.\ \ref{socialflow}d and confirmed by the increased two-step and three-step return rates. This memory effect changes the perceived organization of the scholarly literature. With first-order Markov dynamics, \emph{PNAS} is assigned to a single biological field. With second-order Markov dynamics, however, \emph{PNAS} is assigned to five fields, including cell biology, ecology, and mathematics. Likewise, the multidisciplinary journal \emph{Science} is assigned to ten fields with second-order compared to one field with first-order Markov dynamics. Contrarily, field-specific journals, such as Ecology or Plant Cell, are clustered in single fields both with first- and second-order Markov dynamics. Measured as the average number of module assignments per physical node, we report the module assignments for all networks in Table~\ref{table:stats}. Compared to first-order Markov analysis in the systems analysed here, community detection with second-order Markov dynamics reveals system organizations with more and smaller modules that overlap to a greater extent.

The memory effects on community detection have interesting network-theoretical implications. Community-detection methods typically identify modules with stronger internal than external connections \cite{newman2006modularity,lancichinetti2011finding} or with relatively long flow persistence times \cite{rosvall2008maps,delvenne2010stability}. A problem with these methods is that they tend to assign each node to a very limited number of modules, in contrast to the observation that real modules often show pervasive overlap \cite{ahn2010link,evans2009line,yang2012defining}. Rather than being a shortcoming of the algorithms, our results show that this problem can be a result of distorted modular dynamics in standard networks that prevent the methods from capturing the underlying dynamics and uncovering the actual modules, as with the air traffic example in Fig.\ \ref{memoryairclustering}. Interestingly, some heuristic algorithms for finding highly overlapping modules in standard networks can be seen as trying to account for second-order Markov dynamics (see the Supplementary Information, Sec.\ 3). The clique percolation \cite{palla2005uncovering} and link clustering \cite{ahn2010link} methods are known as topological methods that operate on the network structure without inducing flow on the links. If we take a flow perspective, the percolation of cliques can be seen as restricting flow to stay within connected cliques \cite{palla2005uncovering}. Also, the coupling of links by neighbour similarity can be seen as prolonging flow persistence times in highly connected modules \cite{ahn2010link}. As we show in the Methods section, they are reasonably good at identifying overlapping communities of second-order dynamics aggregated in undirected standard networks. Nevertheless, using empirical data of flow pathways rather than clever assumptions has several advantages. Aggregating links in standard networks inevitably destroys information that cannot be fully recovered. As the benchmark test in Methods shows, a method that operates directly on the flow pathways can achieve superior results.\\

\begin{figure*}[ht]
    \centering
    \includegraphics[width=\textwidth]{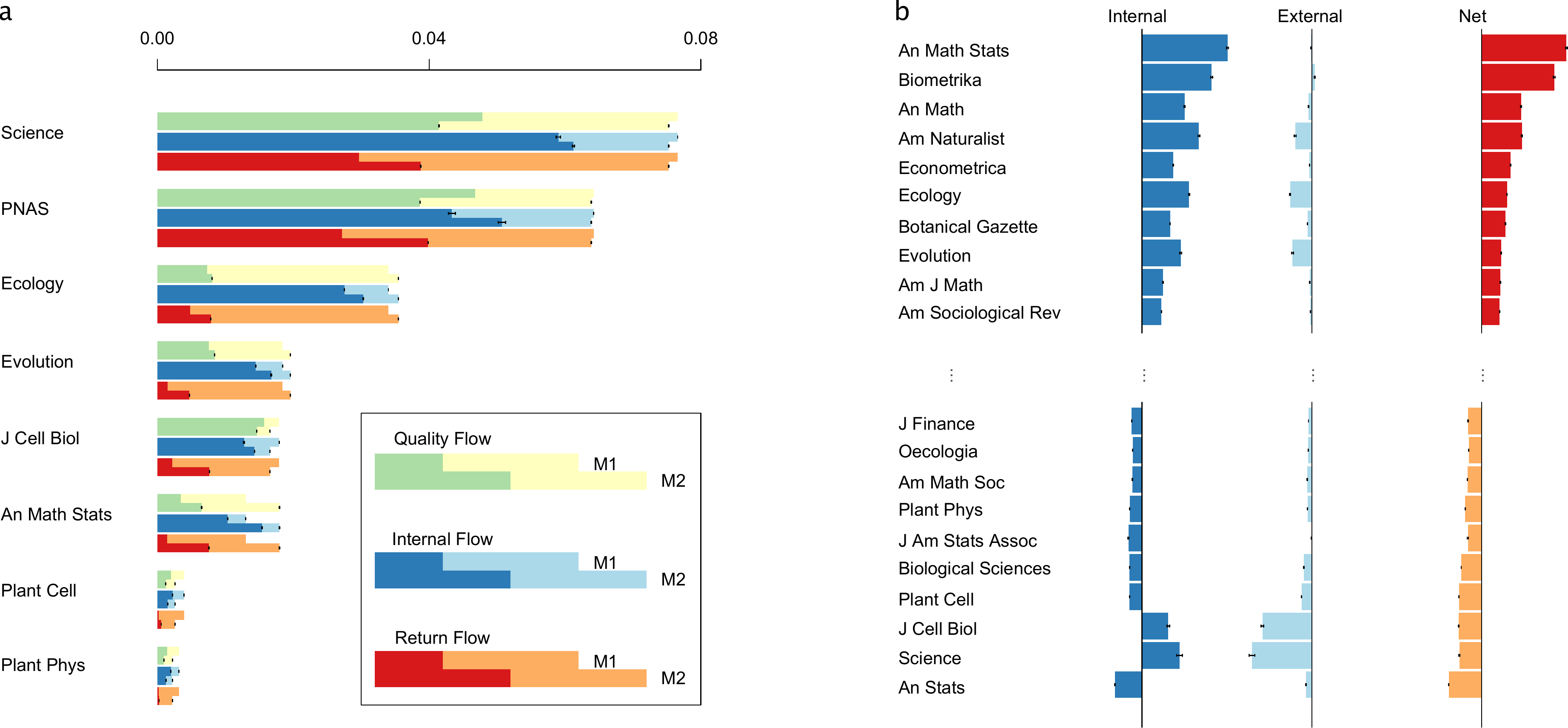}\\
    \caption{\textbf{Memory affects ranking of nodes.} (\textbf{a}) Comparing changes in flow from a first- to a second-order Markov model (M1 to M2). Three kinds of flow were tracked for all journals in both M1 and M2: (1) quality flow from the top ten journals (green) or all other journals (yellow), (2) internal flow coming from journals without crossing community boundaries (dark blue) versus external flow that does cross community boundaries (light blue), and (3) return flow after two steps on the network (dark red) versus lost flow after two steps on the network (light red).  M1 is always the top bar, and M2 is the bottom bar (see legend insert). The error bars indicate the 10th and 90th bootstrap percentiles. The error bars at the intersection of the stacked bars represent the variation in the quality, internal, and return flows, respectively. The error bars at the end of the stacked bars represent the total variation in flow. M1 bars for quality and return flows do not show error bars since the networks are exactly the same. The journals selected for this figure were chosen because they are mentioned in the primary text. (\textbf{b}) Largest gainers and losers in the top 100 journals when including effects of second-order Markov dynamics. The upper portion shows the journals that gain the most in flow, and the lower portion the journals that lose the most in flow. Dark blue indicates a gain/loss in flow coming from journals without crossing community boundaries (internal flow). The light blue indicates a gain/loss in the flow that does cross community boundaries (external flow).  The dark red shows the net gain and orange shows the net loss for each journal listed. The error bars indicate 10th and 90th bootstrap percentiles. \label{flowcomparison}}
\end{figure*}

\noindent\textbf{Memory affects ranking of nodes.} When going from rankings based on counting links to measuring the average visit frequency of a random walker on a standard network, i.e., calculating the PageRank \cite{brin1998anatomy}, the importance of neighbours becomes evident. Similarly, when going to PageRank on a network with second-order memory, the amount of flow received from neighbours also depends on the flow's origin \cite{chierichetti2012web,bohlin2014ranking}.
We define a generalized second-order PageRank as the stationary solution of (\ref{general1})
\begin{equation}
\label{lolo1}
\pi(\vv{jk}) = \sum_{i} \pi(\vv{ij})  p(\vv{ij} \rightarrow \vv{jk}).
\end{equation}
Solving (\ref{general1}) requires finding the dominant eigenvector of the $L \times L$ transition matrix $p(\vv{ij} \rightarrow \vv{jk})$, where $L$ is the number of memory nodes. Note that this matrix is asymmetric even if the original network is undirected, as a transition $\vv{ij} \rightarrow \vv{jk}$ does not exist in the opposite direction $\vv{jk} \rightarrow \vv{ij}$, even if each link is bi-directional. 
After finding $\pi(\vv{jk})$, the centrality of physical nodes in the original network is given simply by
\begin{equation}
\label{lolo2}
\pi(k) = \sum_{j} \pi(\vv{jk}) = \sum_{k} \pi(\vv{jk}),
\end{equation}
where the second equality holds because of conservation of probability (see Methods for details on ergodicity).

In order to illustrate the effect of second-order Markov dynamics on ranking and on PageRank in particular, we focus on the journal citation network (see the Supplementary Information, Sec.\ 4 for analytical results). This example has practical applications because PageRank is a popular measure for ranking the scientific importance of journals \cite{west2010eigenfactor}.  In the citation network, we observe that ten percent of the flow is reallocated when moving from a first-order to a second-order Markov model (see Table~\ref{table:stats}).  Some journals benefit from this reallocation and some do not. The interesting question is: which ones gain and why? 

Figure~\ref{flowcomparison}a shows why some journals increase their ranking from a first- to a second-order Markov model. For example, \textit{Ecology} gains in total flow, which can primarily be explained by the amount of flow coming from high-quality journals (green), the amount of internal flow coming from journals without crossing community boundaries (dark blue), and the amount of flow returning after two steps (dark red). We consider high quality flow to be flow from the top ten journals.  Flow from these journals comprises 1/3 of all flow in the system.  For \textit{Ecology}, there is an increase in return flow and internal flow when moving from a first- to a second-order Markov model, as well as a slight increase in flow from the top ten journals. 

In contrast, the large multidisciplinary journals receive less flow from other top journals. In a first-order Markov model, they \textit{leak} flow between communities and boost each other. For example, \textit{Science} in a first-order Markov model receives flow from and then redistributes flow to journals in multiple fields, even if no readers would cross those field boundaries. In contrast, \textit{Science} in a second-order Markov model mainly receives flow from and redistributes flow to journals within the same fields. Because a significant fraction of the flow that leaks between fields in a first-order Markov model reaches multidisciplinary journals, they receive less flow in a second- relative to a first-order Markov model. As Fig.\ \ref{flowcomparison}b illustrates, journals that increase from a first- to a second-order Markov model almost always see an increase in the flow from their primary community (internal flow). In general, journals that do not depend on leaking flow between modules gain flow, and journals that do, including multidisciplinary journals, lose flow, when two-step memory is taken into account.

We now turn to discussing the advantages of using a second-order Markov model for ranking journals. Since we analyse rankings designed to capture dynamics, the issue with leaking flow of a first-order Markov model directly provides a reason for preferring a second-order Markov model. However, leaking flow is also indirectly associated with another important reason for preferring a second-order Markov model. All rankings are subject to gaming, and a good ranking ought to be difficult to manipulate. For example, the journal impact factor \cite{garfield2006history}, which simply counts the number of citations a journal receives in a given period of time, and corresponds to a zero-order Markov model, can easily be manipulated by editorial policies that encourage self-citations \cite{monastersky2005number}.

A first-order Markov model, in particular one that ignores self-citations \cite{west2010eigenfactor}, is more difficult to exploit, because the value of a citation depends on the ranking of the citing journal. Since important journals need to be cited by important journals, insignificant journals cannot directly boost their own ranking. However, leaking flow is a weak point of this first-order ranking. For example, Fig.\ \ref{socialflow}c illustrates that the first-order citation flows mix and leak from the ecology journals to the molecular biology journals through multidisciplinary \emph{PNAS}. In this way, citations from ecology journals to multidisciplinary journals will indirectly boost molecular biology journals. For improving the ranking of the citing journal, leaking flow therefore creates a potential incentive to reduce the number of citations to multidisciplinary journals. This citation bias works against the principle that citations should go to the best work, and can have have negative influence on the quality of the ranking.

The problem caused by leaking flow is minor for a second-order ranking, since citation flows to multidisciplinary journals tend to return and stay within the citing field. This effect not only explains why multidisciplinary journals lose and field specific journals gain when going from a first- to a second-order model as shown in Fig.~\ref{flowcomparison}b, it also reduces the influence on ranking caused by strategically excluding citations to multidisciplinary journals. For example, while the ranking of \emph{Ecology} improves by removing citations to \emph{Science} and \emph{PNAS} both with a first- and a second-order model, the effect is three times smaller with the second-order model. That is, a second-order Markov model for ranking journals is more robust to manipulation.\\

\noindent\textbf{Memory and spreading processes.} Previous work has considered temporal and memory effects on spreading by modelling time-respecting paths in temporal networks of contacts \cite{rocha2011simulated,lentz2013unfolding,pfitzner2013betweenness} and bidirectional paths in mobility networks of commuters \cite{balcan2011phase,belik2011natural,poletto2012,Keeling11052010}. Our objective is to quantify the full effect of second-order Markov dynamics in general mobility patterns. Therefore, here we model spreading by considering unrestricted second-order Markov processes obtained from empirical pathways.

We considered two classical models for spreading processes \cite{vespignani2012modelling}: a meta-population model that we implemented for the cities and is related to disease spreading, and a simpler model for spreading of ideas or rumours that we studied on the email dataset.
Both models are stochastic compartmental models. In the meta-population model, we use SIR dynamics, and in the simpler model, we use SI dynamics. S, I, and R refer to different categories of individuals: susceptible individuals (S) are healthy individuals who have not been touched by the infection; infected individuals (I) have been reached by the epidemic and in turn can transmit the infection to other individuals; and recovered individuals (R) are those who reach immunization after being infected and cannot spread the disease anymore. 

In the meta-population model, we observe that using a second-order Markov process has a negligible effect on the size of the epidemic, also known as the attack rate, and that it only slightly tends to slow down the spreading process. In contrast, in the simpler model, we observe that second-order Markov dynamics significantly slow down the spreading process.
We conclude that we only observe significant memory effects on the spreading dynamics when the path dependence is preserved at transmission.
For the cities dataset, the effect of second-order Markov dynamics is negligible both because memory is lost at transmission between random individuals in cities, and also because travellers do not return at sufficiently high rate compared to pure commuting traffic \cite{balcan2011phase,belik2011natural,poletto2012,poletto2013human} to limit the number of disease introductions in cities \cite{lessler2009cost}. Below we provide a more detailed discussion.\\

Modelling spreading with SIR dynamics and meta-populations for the cities data set. The model works in two steps, like the reaction-diffusion model proposed in ref.~\citen{colizza2007reaction}. During the reaction step, each infected individual can recover with probability $\mu$ and each susceptible individual can get infected by any infected individual in the same physical node. Effectively, the infection is transmitted regardless of where individuals were one step before and, therefore, describes full mixing at the physical node level. Let us define the total number of individuals in physical node $i$, $P_i$, and the total number of infected individuals in node $i$, $I_i$. We estimated the number of individuals in each city from the number of tickets in our dataset that end in the cities and considered a total population of 300 million, which is a rough estimate of the total population of the US. Assuming that the transmission rate is $\beta/P_i$, where $\beta$ is a parameter that accounts for the virulence of the disease, the probability of each susceptible individual becoming infected is $1 - (1 - \frac{\beta}{P_i})^{I_i}$. The transmission rate is the virulence factor divided by the total population of node $i$, because we assume that each individual can get in touch with a fixed number of other individuals  \cite{colizza2007reaction}. 

After the reaction step, we carry out the diffusion of people in the city network with or without memory of their previous step. Each individual can move to neighbouring cities with probability $\sigma$ if she is ready to start a new trip in a self-memory node, indicating that she was in the same physical node in the previous step, and with probability $1/\tau$ if she is travelling and not in a self-memory node, indicating that she was not in the same physical node in the previous step. We use two different probabilities because the fraction of people who start a new trip from a self-memory node (from home) is much smaller than those who continue the trip after it started. We consider $\sigma=10^{-3}$ days$^{-1}$, which is of the order of magnitude of the number of  new itineraries per day divided by the total population (we estimate $\sigma \simeq 2 \times 10^{-3}$ itineraries per person per day, from our data).
The length of stay $\tau$ can be extremely short if a city is visited just to take a connecting flight. Although the length of stay is heterogeneously distributed \cite{poletto2012}, we simply considered an average length of stay of 2 days. That is, once a trip started, each individual has a 50\% chance of spending another day in the city she is visiting or of moving to the next city. From most memory nodes, it is possible to reach a self-memory node and end the trip, such that the probability of leaving again is $\sigma$.

\begin{figure*}[ht]
    \centering
\includegraphics[width=\textwidth]{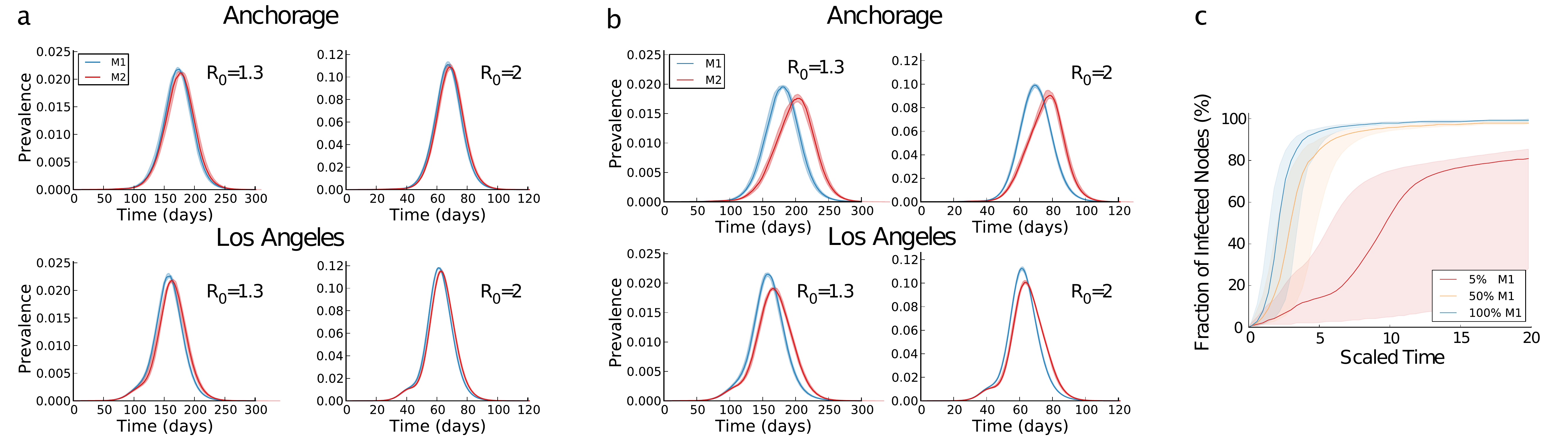}
\caption{\textbf{Memory and spreading processes.} (\textbf{a}) Fraction of infected individuals (prevalence) as a function of time measured in days from the beginning of the process. The two curves represent a first- (M1) and a second-order (M2) Markov process. We seeded the outbreak with $100$ infected individuals in Anchorage (top) and Los Angeles (bottom). (\textbf{b}) Prevalence as a function of time in a modified dataset where people only travel to a city and come back. (\textbf{c}) Fraction of individuals that have received the rumour as a function of scaled time from the beginning of the process. The three curves show different level of mixing between the first- and second-order Markov model. The shaded area gives the $25$ and $75$ percentiles and the solid curve is the median.}
\label{spreadingfig}
\end{figure*} 

After starting a trip, movements can be carried out with a first- or second-order Markov process. Starting from Anchorage and Los Angeles, Fig.\ \ref{spreadingfig}a shows the difference in the evolution of the spreading process. We used $\mu^{-1}=3$ days, and two different values for the basic reproduction number $R_0= \beta\mu^{-1} = $ $1.3$ and $2$, which is the average number of new infections caused by each infected individual before recovering.
The total fraction of infected people at the end of the epidemic is barely affected (the difference is smaller than $10^{-4}$), and there is only a small delay in the spreading process. In order to estimate this delay, we measured the peak time, i.e., the day in which the number of infected individuals is the highest. We averaged the peak times across different runs, with the 100 infected individuals in a particular city selected proportional to its population. For $R_0=1.3$, we estimated the peak time with first-order memory dynamics to $160 \pm 8$ and with second-order memory dynamics to $166 \pm 9$. For $R_0=2$, we estimated the peak time first-order memory dynamics to $62 \pm 3$ and with second-order memory dynamics to $64 \pm 3$. In both cases, the difference is $\simeq 3\%$.

To better understand these results, we repeated the analysis after first removing all but short returning itineraries, such as New York--Chicago--New York. In this way, we can compare with the work on commuting traffic that has reported a slow down in the spreading process \cite{balcan2011phase,belik2011natural,poletto2012,poletto2013human}. For these dynamics, while we still do not observe an effect on the attack rate, Fig.~\ref{spreadingfig}b shows that we observe a significant effect on the peak time by modelling commuting traffic with a second-order Markov process. With only commuting traffic, a second-order Markov model captures that travellers spend only limited amount in other cities, thereby reducing the effective connectivity, and the number of disease introductions in cities. In the actual data, however, the number of one-way tickets and connecting flights is sufficiently large to reduce the return rate and increase the time spent in other cities to a level at which the effect on spreading vanishes between first- and second-order dynamics \cite{lessler2009cost}. Again, once random transmission occurs in a city, all memory effects are washed out in this meta-population model. Therefore, the effect of a higher-order Markov process is primarily influential in the beginning of the outbreak during the introduction phase when the sequence of contacts matters \cite{rocha2011simulated,lentz2013unfolding,pfitzner2013betweenness}. Overall, we conclude that the first- and second-order dynamics must be sufficiently different to show a clear difference on the spreading. To quantify precisely how different is an interesting question for further investigation.\\

Modelling spreading with SI dynamics without meta-populations for the emails data set. In the email data set, each physical node represents an individual with a memory node for each other individual from which an email was received. The target of a memory node's out-link represents the individual to which the email was forwarded to, and the weight the total number of such emails that has been forwarded. We model emails as ``hosts'' for rumours and each individual $j$ can become infected (informed) if she receives an ``infective'' email from an individual $i$. When this happens, memory node $\vv{ij}$ associated with the source becomes infected, and the individual is now informed. The infective email can be forwarded to another person $k$, according to the probability distribution $p(\vv{ij} \to \vv{jk})$. In this way, we model the spread of rumours as a simple contagion process without ``stiflers'' who no longer spread rumours \cite{nekovee2007theory}. Therefore, we focus on the early stages of a spreading process. To study the robustness of the effects of this second-order Markov process, we also allow information to be spread independently of the source at different level of mixing between the first- and second-order Markov model. See Methods for details about the model. 

For this model, we measured the speed of the spreading process. Figure \ref{spreadingfig}c shows the average fraction of individuals that has heard about the rumour as a function of time, starting from a single infected memory node at time $t=0$. The initial nodes were randomly selected among those belonging to the largest strongly connected component. We scaled the time by multiplying by the rumour spreading rate to make results independent of this parameter. Overall, the spreading is much slower when emails are modelled by a Markov model of second order, since this model can capture that most emails are forwarded within strongly confined modules of individuals, which also prevent them from reaching highly connected and efficient spreaders. 
Moreover, the main difference compared with the meta-population model is that an individual informed about a rumour can participate in multiple email conversations simultaneously without an interest in informing everybody about the rumour. That is, where information spreads often depends on from where it is coming.

\section*{\large Discussion}

We have shown that a second-order Markov model is required to capture essential dynamical processes in a variety of integrated systems, with important consequences for community detection, ranking, and information spreading.
Recent work has indicated that a first-order Markov model may fail to adequately predict real dynamics \cite{song2010limits,takaguchi2011predictability,chierichetti2012web,pfitzner2013betweenness}.
That is, real dynamics often have at least one-step memory, which conventional network analysis cannot capture. To bridge this gap, we generalized three commonly used methods of community detection, ranking, and spreading, to operate on a second-order Markov model of flow. We used several real-world and synthetic examples to show that these methods reveal system organizations that better correspond to actual structures, including increased return flow that confines flow in smaller and more overlapping modules. Previously, researchers have tried to reveal such structures with heuristic algorithms, but our approach uses more data rather than extra assumptions, and benchmark and bootstrap analyses show that these results are real and based on sufficient data. Consequently, we have demonstrated that using a second-order Markov model is often essential for fundamental methods in network science.

The combination of our examples indicates that memory is critical for analysing network flows in general, and we expect researchers throughout the sciences to find the methods useful for analysing increasingly available pathway data. Therefore, we have made data and code available online at \href{http://www.icelab.org.umu.se/memorynetworks}{www.icelab.org.umu.se/memorynetworks}, and integrated the community-detection algorithm in the Infomap sofware package available at \href{http://www.mapequation.org}{www.mapequation.org}. Our methods can be directly generalized to higher-order Markov models as well. Even if our statistical analysis of higher-order Markov models suggests that we have captured most of the salient features in the analysed systems, other systems where longer pathway data are relevant and available may have discernible higher-order features. We expect such features to be less salient, and other means of balancing model complexity and utility may be more appropriate.

\section*{\large Acknowledgements} We thank F.~Liljeros for extracting the patient data and JSTOR for providing the journal citation data. We also thank D.~Edler, A.~Ekl\"of, D.~Kolp, C.~Poletto, and D.~Vilhena for many discussions. M.R.\ was supported by the Swedish Research Council grant 2012-3729. R.L.\ was supported by the Belgian Network DYSCO, funded by the IAP Programme initiated by Belspo.

\section*{\large Methods}
\noindent First we provide a short description of how we assembled long real pathways into networks with and without memory. Then we briefly describe how we performed and evaluated the community detection analysis and provide details for how we achieved ergodicity in the ranking analysis. In the Supplementary Information, we detail how we assembled all data, performed the statistical analysis, and developed the community-detection algorithm. We also provide an illustration of how higher-order memory constraints on flow can be modelled without access to full pathways. In addition to its explanatory power, the memory model also summarizes the dynamics of a system with a limited number of parameters and makes it easy to compare the dynamics between different systems.\\

\begin{figure}[ht]
    \centering
    \includegraphics[width=\columnwidth]{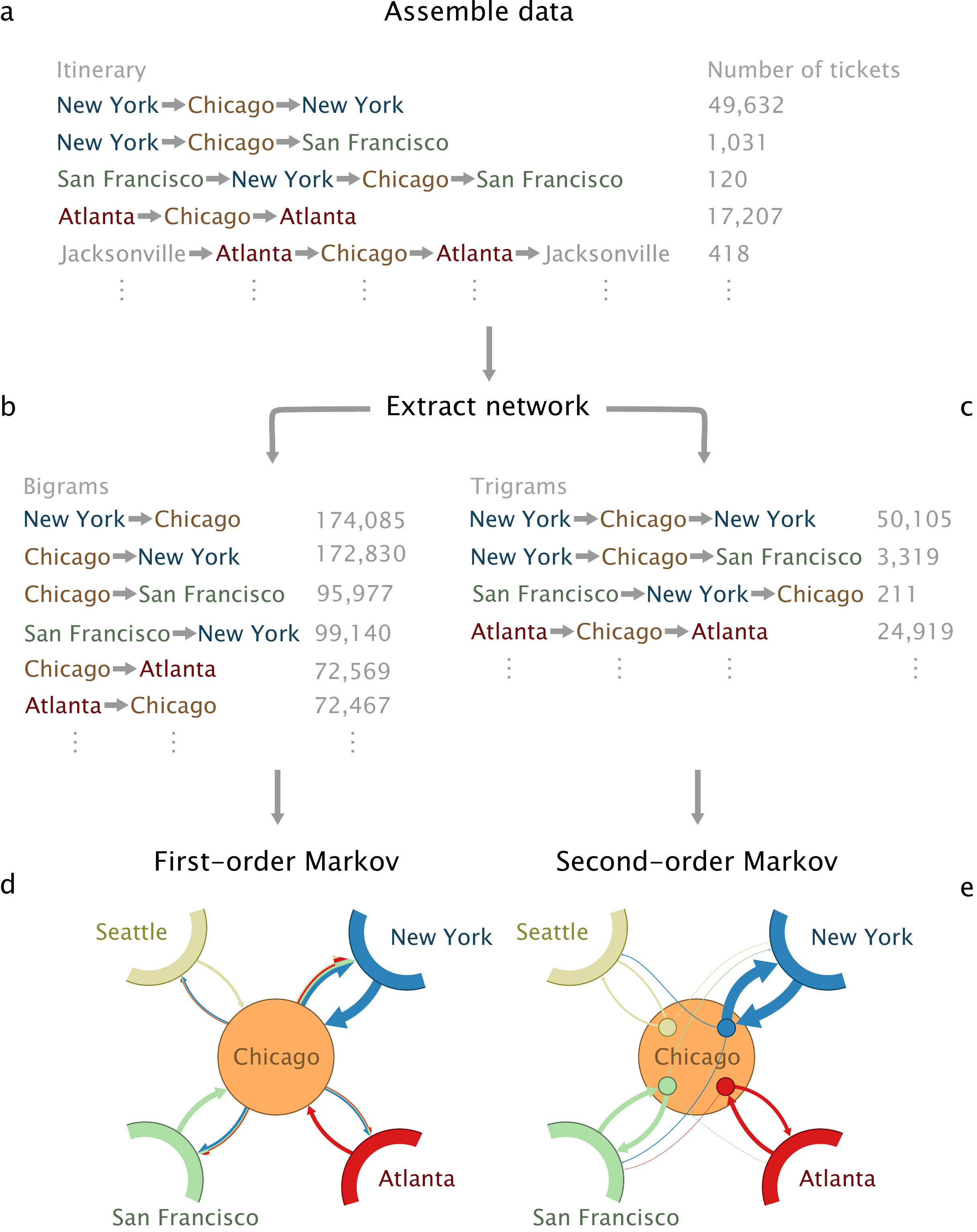}
    \caption{\scriptsize \textbf{From pathway data to networks with and without memory.} (\textbf{a}) Itineraries weighted by passenger number.  (\textbf{b}) Aggregated bigrams for links between physical nodes. (\textbf{c}) Aggregated trigrams for links between memory nodes. (\textbf{d}) Network without memory. (\textbf{e}) Network with memory. Corresponding dynamics in Figs.\ \ref{socialflow}\textbf{a} and \textbf{b}.\label{fromdatatonetworks}}
\end{figure}

\noindent\textbf{Assembling pathways into networks with and without memory.}
Figure \ref{fromdatatonetworks} illustrates how we generated the networks that describe the dynamics in Figs.\ \ref{socialflow}a and b: from pathways in a, via weighted links in b and c, to directed weighted networks in d and e. First we collect long pathways, in this example, of real itineraries from The Research and Innovative Technology Administration (RITA) (Fig.\ \ref{fromdatatonetworks}a). The data contain each stop on 19,415,369 itineraries with average pathlength 3.3 between 464 airports in the US. We used data from the first three quarters of 2011, which contain a sample of 10\% of all itineraries during the time period. In the cities data set, we aggregated all airports within a radius of 50 kilometres and called destinations by corresponding city names. Each pathway has a weight equal to the number of passengers who have purchased exactly that itinerary. To generate weighted directed links for the standard network, we counted bigrams (city pairs) in the itineraries (Fig.\ \ref{fromdatatonetworks}b). To generate weighted directed links for the memory network, we counted trigrams (city triplets) in the itineraries (Fig.\ \ref{fromdatatonetworks}c). In the airport data set, we focused on transfer traffic and disregarded one-way trips with a single flight (21\% of all itineraries). In the cities data set, however, we focused on real passenger traffic for accurate modelling of disease spread and included also short pathways. Therefore, in the cities data set, the typical memory averaged over all travellers is somewhat less than second order. Then we assembled the links into networks. All links with the same start node in the bigrams represent out-links of the start node in the standard network (Fig.\ \ref{fromdatatonetworks}d). A physical node in the memory network, which corresponds to a regular node in a standard network, has one memory node for each in-link (Fig.\ \ref{fromdatatonetworks}e). A memory node represents a pair of nodes in the trigrams. For example, the blue memory node in Fig.\ \ref{fromdatatonetworks}e represents passengers who come to Chicago from New York. All links with the same start memory node in the trigrams represent out-links of the start memory node in the memory network. In this way, the memory network can maintain dependency between where passengers come from and where they go next. Figures \ref{socialflow}a and b show the dramatic effect of maintaining second-order memory: passenger travel is much more constrained than what the standard network can capture. See the Supplementary Information, Sec.\ 1, for details of how we obtained pathways for all analysed networks and represented them as networks with and without memory.\\

\noindent\textbf{Significance analysis with resampling.} We performed two different statistical tests to validate our results, bootstrap resampling of all summary statistics in Table 1 and surrogate data testing of the Markov order in Fig.\ \ref{markovtest} and Supplementary Fig.\ 3. Bootstraping allows us to assign confidence intervals to the summary statistics based on resampling of the observed dataset. Accordingly, only trigrams observed in the data will occur, but possibly with different frequencies. Contrarily, surrogate data testing allows us to also generate unobserved trigrams and is therefore suitable for hypothesis testing of the Markov order against a null model. In turn, we describe the two methods below.

For the bootstrapping, we generated 100 bootstrap replicas for each dataset by resampling the weights of the pathways from a multinomial distribution (for patients, taxis, and emails, we only had access to trigrams and resampled their weights directly). This scheme corresponds to resampling of all pathways with replacement. That is, we assume that pathways are generated independently. For the air traffic depicted in Fig.\ \ref{fromdatatonetworks}a, for example, we assume that tickets are bought independently. This assumption of independence is, of course, only approximately true, but since flight tickets rarely are bought for more than a few passengers at the same time, the approximation will work well in practice. After resampling the pathways, we generated the networks as described in Figs.\ \ref{fromdatatonetworks}b-e and performed any analysis as on the raw network. For each set of summary statistics, we calculated the bootstrap confidence interval by ordering the 100 bootstrap estimates and eliminated the ten smallest and ten largest estimates. In general, we report the lower and upper limits of this interval.

For the surrogate data testing, our null hypothesis was that the flow is first-order Markov, and we used the conditional entropy at each node as a test statistic. Assuming that the null hypothesis is true, we estimated the probability that the conditional entropy in a second-order Markov process is at least as low as the observed value. We estimated this probability, the p-value, with surrogate resampling and rejected the null hypothesis if the p-value was lower than 0.10. For each node and for each resampling, we removed the second-order Markov effect by performing random pairings between all nodes visited before and after the node given by all trigrams centred at the node. With this resampling scheme, we can single out nodes with a significant second-order Markov effect. See the Supplementary Information, Sec.\ 2, for further details and for surrogate testing of higher Markov orders.\\

\noindent\textbf{Community detection with second-order Markov dynamics.} We have chosen to work with the flow-based map equation framework \cite{rosvall2008maps}. In principle, we could have used alternative flow-based methods \cite{delvenne2010stability}, but the map equation framework allows us to compare the community structure with first- and second-order Markov dynamics by only modifying the dynamics and not the mechanics of the method. Since we are interested in overlapping modules, we build our new method on a generalization of the map equation to overlapping modules \cite{esquivel2011compression}.
\newline \indent
The map equation framework is an information-theoretic approach that takes advantage of the duality between compressing data and finding regularities in the data. Given module assignments of all nodes in the network, the map equation measures the description length of a random walker that moves from node to node by following the links between the nodes. Therefore, finding the optimal partition or cover of the network corresponds to testing different node assignments and picking the one that minimizes the description length \cite{rosvall2008maps}.

The map equation framework easily generalizes to higher-order Markov dynamics, because memory networks only change the dynamics of the random walker as described above. Therefore, instead of applying the search algorithm on the standard network, we apply it on the memory network and assign memory nodes to modules, with one important difference: Since we are interested in movements with or without memory between physical nodes, the description of the random walker must reflect this process. Therefore, when two or more memory nodes of the same physical node are assigned to the same module, the description length must capture the fact that the memory nodes share the same codeword. We achieve this description by summing the visit frequencies of all memory nodes of each physical node in a module and then use this visit frequency to derive the optimal codeword length. We ensure that the community detection results only depend on memory effects by representing first-order Markov dynamics in a memory network, with each memory node having the out-links of its corresponding physical node in the standard network. In this way, the compression algorithm remains the same and only the dynamics change.

\begin{figure}[ht]
    \centering
    \includegraphics[width=\columnwidth]{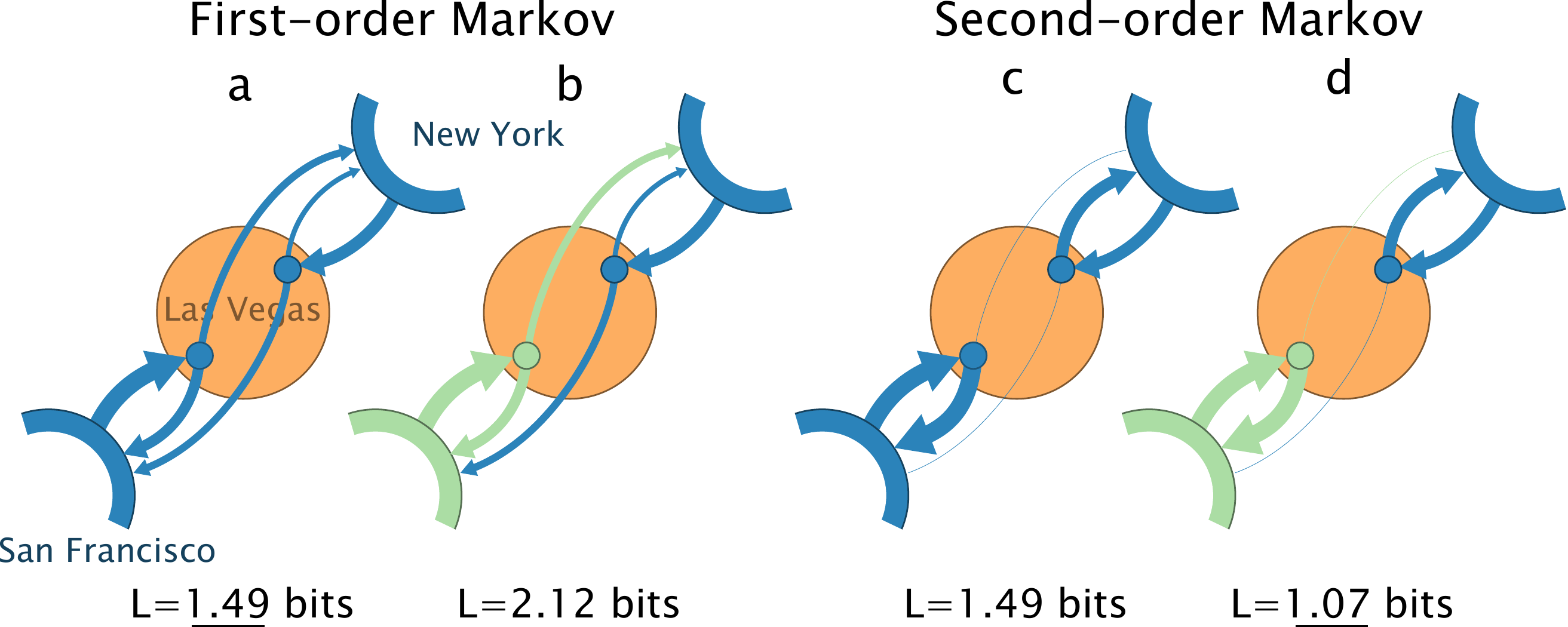}
    \caption{\scriptsize \textbf{Second-order memory dynamics reveal overlapping modules.} Pathway data between San Francisco, Las Vegas, and New York, represented with memory nodes capturing first-order (\textbf{a}-\textbf{b}) and second-order (\textbf{c}-\textbf{d}) memory dynamics. With first-order memory, the characteristic out-and-back travel of Las Vegas is lost and the dynamics are best described as movements in one module; describing the dynamics with two overlapping modules requires 0.63 more bits. With second-order memory, the out-and-back travel is evident and the dynamics are best described as movements in two overlapping modules, since movements between the modules are very rare. See Supplementary Fig.\ 4 for a detailed derivation of the description lengths.\label{memorynetworkclustering}}
\end{figure}

Figure \ref{memorynetworkclustering} illustrates the effect of second-order Markov dynamics on community detection. The pathways represent air travel between San Francisco, Las Vegas, and New York, and correspond to a subset of the itineraries in the city data. With first-order Markov dynamics, there are no regularities to take advantage of in a modular description, and clustering all the cities together gives a shorter description length. With second-order Markov dynamics, however, the strong out-and-back travel pattern to and from Las Vegas makes it more efficient to describe the dynamics as two overlapping modules, with Las Vegas assigned to both modules. That is, the first-order dynamics obscure the actual travel pattern and prevent a modular description from compressing the data. See the Supplementary Information, Sec.\ 3, for further details.

To validate our method, we have performed benchmark tests on synthetic pathways.
We first describe how we build artificial pathways such that flow tends to stay inside predefined communities when described by a second-order Markov model. Then we show that Infomap for memory networks, the community-detection algorithm we have developed, can recover the planted structure. However, when the artificial pathways are described by a first-order Markov model in a standard network, much of the structure is washed out. We show that neither Infomap nor other commonly used methods for overlapping communities can accurately recover the planted structure.

We used the following algorithm to generate trigrams within and between communities:

As planted structure, we consider 128 nodes and the community size fixed to 32 nodes, like in the Girvan-Newman benchmark \cite{girvan2002community}. Moreover, we tune the number of communities $M$. If $M=4$, each node is assigned to a single community. If $M>4$, multiple memberships are assigned to nodes in random order, with the constraint that no node can be assigned to the same community twice.

As synthetic pathways, we draw $E_{in}$ internal trigrams and $E_{out}$ external trigrams. Internal trigrams are paths of three nodes $i,j,k$ such that if nodes $i$ and $j$ belong to community $C$, node $k$ also belongs to $C$. For external trigrams, at least two of the three nodes are not assigned to the same community. Below we describe a simple sampling algorithm. In these tests, we set $E_{in}=50,000$, and $E_{out}=5,000$ and $20,000$, respectively. The number of trigrams is relatively high compared to the network size, because to highlight the effect of memory the number of trigrams must be of the same order of magnitude as the number of memory nodes ($128 \times 127 \simeq 15,000$). 
   
Internal trigrams confine flow inside communities. Therefore, if the flow goes from node $i$ to $j$ in community $C$, the next node $k$ must also belong to community $C$. This constraint requires that memory nodes $\vv{ij}$ and $\vv{jk}$ are uniquely assigned to community $C$, although physical nodes $i$, $j$ and $k$ can have multiple memberships. We assign memberships to memory nodes and draw internal trigrams in the following way: 
\begin{itemize}
  \item We uniformly select a community $C$
  \item We uniformly sample nodes $i$, $j$ and $k$ assigned to community $C$. Since nodes can be drawn from multiple clusters, we check that neither memory node $\vv{ij}$ nor memory node $\vv{jk}$ has been assigned to a community different from $C$ yet. If at least one has been assigned to a different cluster, we sample new nodes. If not, we assign the memory nodes to $C$ and record the trigram $i,j,k$.
\end{itemize}

External trigrams guide flow between communities. Therefore, we draw random trigrams $i,j,k$ until at least two of the three nodes have no memberships in common.

\begin{figure}[ht]
 \centering
 \includegraphics[width=1.0\columnwidth]{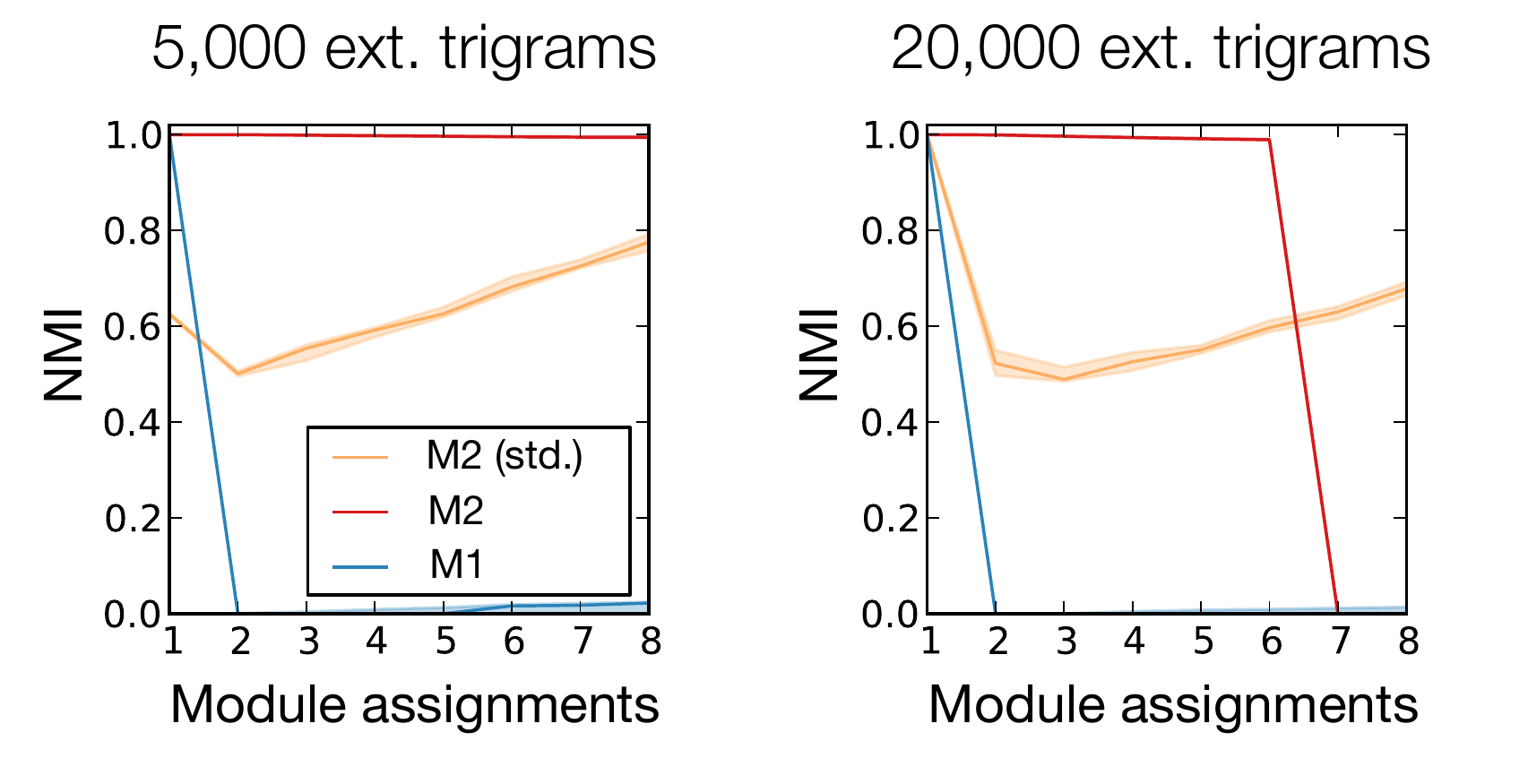}
 \caption{\scriptsize \textbf{Performance tests on benchmark networks.} Performance of Infomap. The blue and red curves refer to M1 and M2 structural information, respectively, whereas the yellow curve was obtained by running standard (undirected) Infomap on a network in which each memory node is treated as a physical node. Lines show median values and shaded areas cover $25$ and $75$ percentiles.}
 \label{fig:M2bench}
\end{figure}

To measure how well Infomap for memory networks recovers the planted structure, we applied the Normalized Mutual Information (NMI) described in ref.~\citen{danon2005comparing} to the community assignments of the memory nodes (we used $\max$ function for the normalization instead of the average). Some memory nodes were only sampled in external trigrams and not assigned a membership by the algorithm above. Since these nodes are not present in the planted structure, we also discard them in the output of Infomap. Figure~\ref{fig:M2bench} shows the performance of Infomap for memory networks with first- and second order Markov dynamics, as well as the performance of standard (undirected) Infomap \cite{rosvall2008maps} with all memory nodes treated as physical nodes. Infomap for memory networks recovers the planted partitions almost perfectly up to at least eight community assignments per node with 5,000 external trigrams and up to six community assignments per node with 20,000 external trigrams. However, with first-order dynamics, Infomap for memory networks is only able to recover the correct partition when no overlap is present. Quite the opposite, standard Infomap tends to find many more modules because the algorithm considers each memory node to be ``independent,'' and there is no intrinsic compression gain from clustering memory nodes of the same physical node together. 

To demonstrate that second-order Markov information is necessary, we aggregated the trigrams into standard undirected networks and applied several commonly used algorithms for overlapping communities. Since the nodes can be assigned to multiple communities, we used the definition of NMI proposed in ref.~\citen{lancichinetti2009detecting} for all methods except for the link clustering method. This algorithm returns a partition of non-overlapping links, which we treated as memory nodes and computed the NMI as described for Infomap above. Since the link clustering method only accepts unweighted graphs as input, we used a threshold of $12$ for link weights and $0.7$ for selecting a partition from the dendrogram, and found that results are not sensitive to these choices. Further, the clique percolation method was unusably slow with all links included and we had to remove links with weights below a certain threshold. We used a threshold of $3$ for $5,000$ external trigrams and $8$ for $20,000$ external trigrams. We also had to provide the clique size ($\simeq 30$).

\begin{table}[ht]
\caption{\textbf{Performance test for overlapping communities on aggregated benchmark networks}\label{table:benchmark}}\vspace{-1em}
\flushleft
\setlength{\tabcolsep}{2.9pt}
\setstretch{0.85}
\begin{footnotesize}
\begin{tabular}{lclllclllclll}
\mytoprule\noalign{\smallskip}
    External trigrams & & \multicolumn{3}{l}{0} &\rule{0.5em}{0em}& \multicolumn{3}{l}{5,000} &\rule{0.5em}{0em}& \multicolumn{3}{l}{20,000}\\
    \noalign{\smallskip}
    Module assignm. & & 1 & 2 & 3 & & 1 & 2 & 3 & & 1 & 2 & 3 \\ \noalign{\smallskip}
    \cline{0-0}\cline{3-5}\cline{7-9}\cline{11-13}\noalign{\smallskip}
    Clique perc. \cite{palla2005uncovering} &&
    $1.0$    &    $1.0$    &   $1.0$ && 
    $1.0$    &    $1.0$    &   $?$ && 
    $1.0$    &    $1.0$    &   $?$ \\
    COPRA \cite{gregory2010finding} &&
    $1.0$ & $0.03$ & $-$ && 
    $1.0$ & $-$ & $-$ && 
    $1.0$ & $-$ & $-$ \\
    Link clust. \cite{ahn2010link} &&
    $1.0$  &  $0.42$  &  $0.32$ &&
    $1.0$  &  $0.43$  &  $0.30$ && 
    $1.0$  &  $0.42$  &  $0.25$ \\
    MOSES \cite{mcdaid2010detecting} &&
    $1.0$ & $1.0 $ & $-$ && 
    $-$ &  $-$ & $-$ && 
    $-$ &  $-$ & $-$ \\
    OSLOM \cite{lancichinetti2011finding} &&
    $1.0$ & $0.97$  & $0.8$ && 
    $1.0$ & $0.80$  & $-$ && 
    $1.0$ & $0.90$  & $-$ \\
    \noalign{\smallskip}
\mybottomrule\noalign{\smallskip}
\end{tabular}
\parbox{\columnwidth}{\scriptsize\raggedright Score measured as the average Normalized Mutual Information.
A dash indicates that the algorithm returned a single module or 128 modules with single nodes. A question mark indicates that the algorithm did not finish.}
\end{footnotesize}
\end{table}

Table~\ref{table:benchmark} shows the results. The clique percolation method was the only algorithm that was able to recover the correct partition with external trigrams and multiple community assignments. However, regardless of the thresholds we tried, for more than $2$ community assignments per node we were not able to obtain any result after several days of running time. The reason why the algorithm is successful on this benchmark test, at least in theory, is that the number of trigrams is so high that the planted communities are cliques of 32 nodes. Of all tested algorithms, the link clustering method was the only one that obtained non-trivial solutions for three or more community assignments per node. In the next section, we illustrate how clique percolation and link clustering can identify overlapping communities of second-order dynamics aggregated in standard networks.\\

\noindent\textbf{Ergodic second-order Markov dynamics.} The solution of (\ref{lolo1}) is not well-defined when the process is not ergodic, which happens when the memory network is not strongly connected, or when it contains closed cycles  \cite{langville2004deeper}.
To circumvent this limitation and to ensure the ergodicity of the stochastic process, we perform two modifications.
First, if a memory node is a dangling node and has no out-links, we use M1 data and assign all out-links from the physical node to the dangling node. In this way, link weights and M1 data become our fallbacks when there is not enough M2 data for an ergodic process on the memory network.
Second, it is standard to allow walkers to randomly teleport across the system, as we mentioned before. Walkers either follow links with probability $\alpha$ or teleport with probability $1-\alpha$ \cite{brin1998anatomy}. 
Therefore, the PageRank of a memory node is given by
\begin{eqnarray}
\label{teleportation}
\label{ergodicsolution}
P(\vv{jk};t+1) &=& \alpha \sum_{i} P(\vv{ij};t)  p(\vv{ij} \rightarrow \vv{jk}) \cr
&+& (1-\alpha) \frac{\sum_j W(j \to k)}{\sum_{lm} W(l \to m)}.
\end{eqnarray}
It is important to note that walkers do not teleport uniformly to memory nodes, but at a rate proportional to the weight $W$ of the corresponding link. Equivalently, walkers thus teleport to physical nodes at a rate proportional to their in-strength. This choice is motivated by recent research showing that so-called {\em link teleportation} improves robustness of ranking with respect to standard teleportation  \cite{lambiotte2012ranking}. A random walk with teleportation is ergodic for any $\alpha<1$, whatever the topology of the underlying network, and its stationary solution can be found by using standard iteration methods.

The link-teleportation scheme works well for ranking nodes, but further improvements can be made for the map equation, which also explicitly operates on the flow between nodes. For community-detection results that are more robust to the particular choice of teleportation parameter, we do not use teleportation steps between nodes and only steps along links to derive the optimal codeword lengths. We achieve the same PageRank of memory nodes in (\ref{ergodicsolution}) by first calculating the stationary distribution with recorded teleportation to physical nodes at a rate proportional to their \emph{out-strength}, followed by a subsequent recorded step without teleportation. By only encoding the last step in this smart teleportation scheme \cite{lambiotte2012ranking}, the community detection results are based on the same ergodic node visit rates as in (\ref{ergodicsolution}), but without the noise on links caused by random teleportation.\\

\noindent\textbf{SI dynamics on networks with memory.} Here we describe how we model spreading with SI dynamics without meta-populations for the emails data set. We assume that each memory node $\vv{ij}$ forwards $ \phi \times s_{\vv{ij}}$ emails per time-step, where $\phi$ is a proportionality constant and $s_{\vv{ij}}$ is the out-strength of the memory node, i.e., the sum of the weights of the links $\vv{ij} \to \vv{jk}$.
A forwarded email from memory node $\vv{ij}$ goes to a memory node, say $\vv{jk}$, with probability $p(\vv{ij} \to \vv{jk})$. If $\vv{ij}$ is informed and $\vv{jk}$ is not, we assume that an email from $\vv{ij}$ to $\vv{jk}$ informs $\vv{jk}$ with probability $\beta$, the so-called rumour spreading rate. 
Let $\tau(\vv{ij} \to \vv{jk})$ denote the overall probability that an infected memory node $\vv{ij}$ transmits the rumour to an uninformed memory node $\vv{jk}$.
Since the probability that the infection is \emph{not} transmitted is the probability that each email leaving $\vv{ij}$ either is forwarded to a memory node other than $\vv{jk}$ or is forwarded to $\vv{jk}$ but ignored, we have:
\begin{align}
 1- \tau(\vv{ij} \to \vv{jk}) &= (1- \beta p(\vv{ij} \to \vv{jk}) )^{\phi s_{\vv{ij}}}\\
 &\simeq e^{-\beta \phi W(\vv{ij} \to \vv{jk}) },\nonumber
\end{align}
where we assume that $\beta$ is small and $W(\vv{ij} \to \vv{jk})=s_{\vv{ij} } \,\, p(\vv{ij} \to \vv{jk})$ is simply the weight of link $\vv{ij} \to \vv{jk}$  in the memory network. In this limit the only relevant parameter is thus $\beta \phi$. Without loss of generality, we can set $\phi=1$, in which case the dynamics of the spreading process are driven by
\begin{equation}
 \tau(\vv{ij} \to \vv{jk})= 1-  e^{-\beta  W(\vv{ij} \to \vv{jk}) }.
 \label{eq_simodel}
\end{equation}
This equation shows that this spreading process with second-order Markov dynamics corresponds to traditional spreading models but performed on memory nodes. That is, the only differences are that emails are forwarded to the next destination depending on where they come from.
Since rumours not necessarily need to spread between individuals that participate in the same email conversations, we allow each informed individual to send emails according to a first-order Markov model with probability $\eta$ at each time step.
Therefore, to study the effects of memory on this spreading process, we can simply tune $\eta$. For example, the extreme case $\eta=100\%$ corresponds to a first-order Markov model. 


\clearpage

\renewcommand{\figurename}{Figure}
\renewcommand{\thefigure}{S\arabic{figure}}
\renewcommand{\tablename}{Table}
\renewcommand{\thetable}{S\arabic{table}}
\renewcommand{\theequation}{S\arabic{equation}}

\renewcommand{\thesection}{\arabic{section}}
\renewcommand{\thesubsection}{\thesection.\arabic{subsection}}
\renewcommand{\thesubsubsection}{\thesubsection.\arabic{subsubsection}}

\setcounter{figure}{0}
\setcounter{table}{0}

\section*{\LARGE Supplementary Information}

\section{Data acquisition and processing}
Here we describe how we estimate the transition probabilities from empirical data. We use empirical data that consist of large sets of multi-step pathways of varying length. To create memory networks that capture an nth-order Markov process, we count all pathways of length $n+1$ in the empirical pathways of length $n+1$ or longer. Take, as an example, the three pathways
\begin{align}\label{pathways}
i \to j \to k \to l,\text{ }l \to k \to j \to i,\text{ and }j \to k \to j.
\end{align}

In a standard network with physical nodes $i$, $j$, $k$, and $l$, we extract the four directed links $i \to j$, $j \to i$, $k \to l$, and $l \to k$ with weight 1 and the two directed links $j \to k$ and $k \to j$ with weight 2, as shown in Supplementary Fig.\hs\ref{memnetschematic}a. These links are all pathways of length 1 that can be extracted from the three pathways in Supplementary Equation (\ref{pathways}). The weight $W(i \to j)$ of a link corresponds to the number of times the link occurs in the pathways. With $W(i \to j) = 0$ if there is no link from $i$ to $j$, the transition probabilities take the form
\begin{align}
p(i \to j) = \frac{W(i \to j)}{\sum_k W(i \to k)}.
\end{align}

\begin{figure}[ht]
    \centering
   \includegraphics[width=\columnwidth]{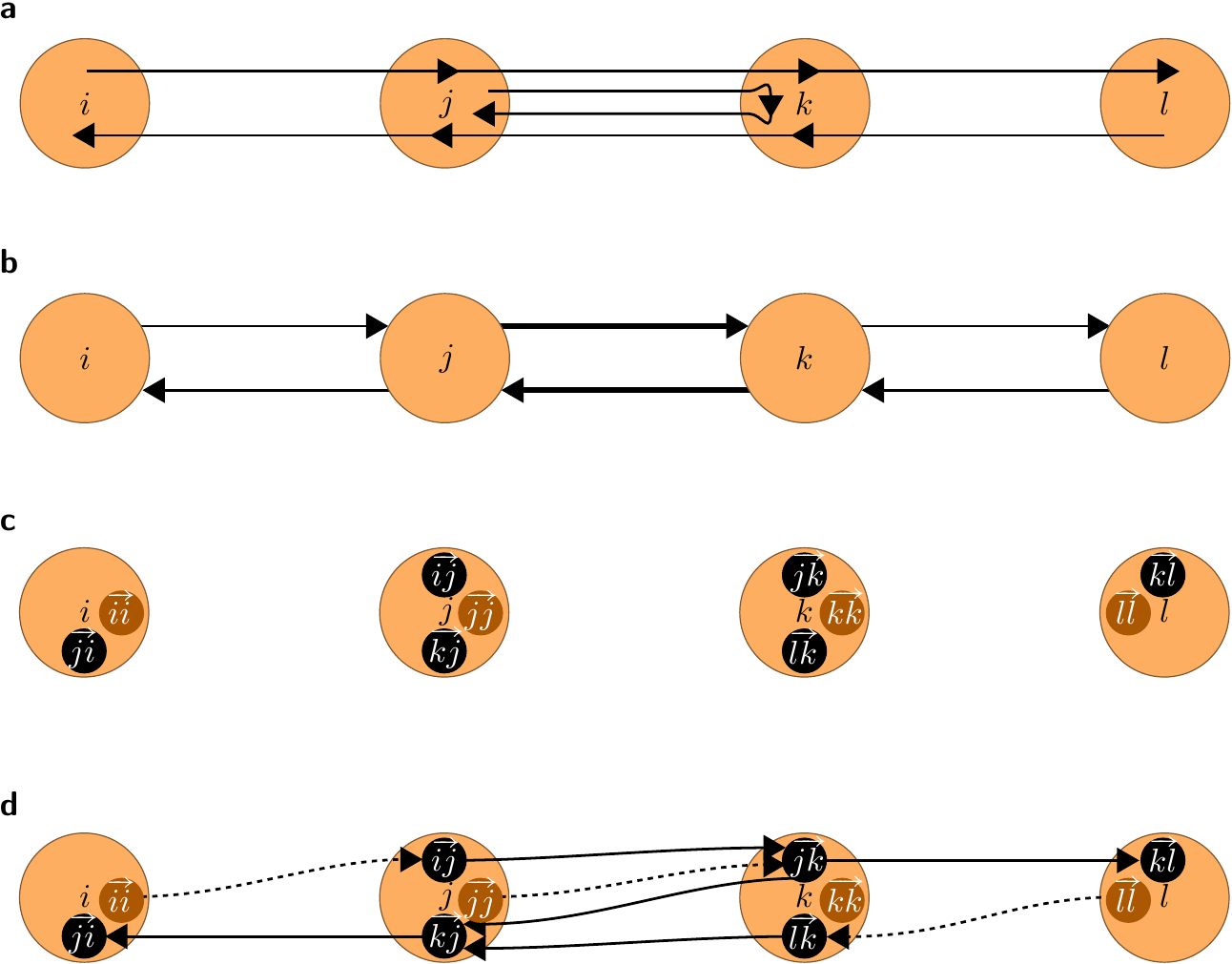}\\
    \caption{\textbf{Representing pathways with a standard network and a memory network.} (\textbf{a}) The three pathways between four physical nodes in Supplementary Equation (\ref{pathways}). (\textbf{b}) Standard network representation. (\textbf{c}) Memory network with memory nodes in black and self-memory nodes in brown. (\textbf{d}) Memory network representation. Dashed links in \textbf{d} from the self-memory nodes represent the first step of each pathway in \textbf{a}. We only use self-memory nodes in the cities dataset (see Sec.\ 1). For the other networks, we only include the first step of each pathway in the weights of the teleportation scheme as described in the Methods section of the main text. \label{memnetschematic}}
\end{figure}

Instead, in a memory network with memory nodes $\vv{ij}$, $\vv{jk}$, $\vv{kl}$, $\vv{lk}$, $\vv{kj}$, and $\vv{ji}$, we extract the five directed links $\vv{ij} \to \vv{jk}$, $\vv{jk} \to \vv{kl}$, $\vv{lk} \to \vv{kj}$, $\vv{kj} \to \vv{ji}$, and $\vv{jk} \to \vv{kj}$, all of weight 1 in this case, as shown in Supplementary Fig.\hs\ref{memnetschematic}d. These links are all pathways of length 2 that can be extracted from the three pathways in Supplementary Equation (\ref{pathways}). With $W(\vv{ij} \to \vv{jk})$ to denote the weight of a link $\vv{ij} \to \vv{jk}$ and $W(\vv{ij} \to \vv{jk}) = 0$ if there is no link from $\vv{ij}$ to $\vv{jk}$, the transition probabilities take the form
\begin{align}
p(\vv{ij} \to \vv{jk}) = \frac{W(\vv{ij} \to \vv{jk})}{\sum_l W(\vv{ij} \to \vv{jl})}.
\end{align}

This formalism can be extended to memory networks that capture even higher-order Markov processes, but here we focus on describing memory networks that capture second-order Markov processes. In this case, links between physical nodes in the standard network representation take the role of memory nodes in the memory network representation. Therefore, a memory network is a form of line graph, but we use the term memory network to highlight our purpose with this representation: to capture movements between physical nodes with transition rates that depend on the past. In network science, line graphs have recently been introduced for a different purpose, namely, to move the focus from nodes to links as a computational trick to detect overlapping modules in networks \cite{SI-evans2009line,SI-ahn2010link}. Supplementary Figure\hs\ref{memnetschematic} shows that the memory network contains more information than a line graph derived from the standard network would. Memory node $\vv{ij}$, for example, corresponds to the link $i \to j$ in the standard network representation. However, this link is not connected with both link $j \to k$ and link $j \to i$, as the standard network suggests, but only with link $j \to k$, as the memory network shows. 

With very long empirical pathways that generate ergodic processes on the memory networks, we could directly use (7) in the main manuscript without any extra work. In practice, however, most pathways are between three and six steps long and boundary effects can influence the analysis. Also, as with directed standard networks, a Markov process on a memory network is rarely ergodic without introducing a small teleportation probability as described in the Methods section of the main manuscript.

Since we are interested in second-order Markov dynamics, it is convenient to store all pathway data as trigrams. From the pathways in Supplementary Equation (\ref{pathways}), we store the trigrams in the following way:\\
\begin{minipage}{\textwidth}
\begin{verbatim}

  i i j 1
  i j k 1
  j k l 1
  l l k 1 
  l k j 1
  k j i 1
  j j k 1
  j k j 1

\end{verbatim}
\end{minipage}
The last number in each line gives the frequency of the corresponding trigram. It is straightforward to calculate the transition probability (6) in the main manuscript between memory nodes from the list of trigrams. For instance, $p(\vv{ij} \rightarrow \vv{jk}) = 1$, $p(\vv{jk} \rightarrow \vv{kl}) = 1/2$, etc.

To be able to initiate dynamics in physical nodes, we repeat the first physical node of pathways twice. We call the corresponding memory nodes self-memory nodes. They represent flow in a physical node that was in the same physical node in the previous step. For example, the first trigram of the pathway $i \to j \to k \to l$ is \verb+i i j 1+ with added self-memory node $\vv{ii}$ (the three trigrams with self-memory nodes above correspond to the dashed links in Supplementary Fig.\hs\ref{memnetschematic}d). In this way, the second pair of nodes in each line forms the link between physical nodes in a standard network representation and it becomes straightforward to obtain M1 data. The procedure is to simply ignore the first node in the trigrams and, for each link between two physical nodes $i$ and $j$, aggregate all weights to $W(i \to j)$. We stress that we use the self-memory nodes only to obtain link weights and M1 data for teleportation steps and not for regular steps (the city network is an exception; see Sec.\ 1). 

When parsing the data, the first line of the trigram data reads literally: ``\emph{There is one link between self-memory node $\vv{ii}$ and memory node $\vv{ij}$}.'' The meaning is: ``\emph{One pathway starts in $i$ and continues to $j$},'' and we increase the weight of memory node $\vv{ij}$ by 1. The second line of the trigram data reads literally: ``\emph{There is one link between memory node $\vv{ij}$ and memory node $\vv{jk}$}.'' The meaning is: ``\emph{One pathway in $j$ came from $i$ and continues to $k$},'' and we add a link from memory node $\vv{ij}$ to memory node $\vv{jk}$ and increase the weight of memory node $\vv{jk}$ by 1.

\subsection*{Cities and Airports} 
\label{sec:city_data}

We compiled the airline pathways from the Airline Origin and Destination Survey (DB1B) made public by the Research and Innovative Technology Administration (RITA). The data contain each stop on 19,415,369 itineraries between 464 airports in the US with average pathlength 3.3 (21\% of length two, 53\% of length three,  5\% of length four, 19\% of length five, 1.4\% of length six, and 1.0\% of length seven or longer). Note that the origin is included in the path length, such that a path length of three corresponds to two flight legs. We used data from the first three quarters of 2011.

In the city memory network, we aggregated all airports within a radius of 50 kilometres. Since the airport data have clear starts and stops --- a passenger is based in a city and most often returns to the same city at the end of the itinerary --- we also include the home city itself in the analysis. In this way, we can better capture real passenger traffic and extend the analysis beyond transfer traffic. We thus represent the home city with the corresponding self-memory node in the memory network and, unlike in the other memory networks, include the self-memory nodes in the analysis. For example, we represent a pathway $i\ldots j\ldots k \ldots j \ldots i$ going from city $i$ to city $k$ with transfer in city $j$ and back with the trigrams\\
\begin{minipage}{\textwidth}
\begin{verbatim}

  i i j 1
  i j k 1
  j k j 1
  k j i 1 
  j i i 1

\end{verbatim}
\end{minipage}
and include the first and last trigrams in the analysis.

In the airport memory network, we focused on transfer traffic and did not include self-memory nodes. For example, a pathway $i\ldots j\ldots k \ldots j \ldots i$, representing an itinerary from airport $i$ to airport $k$ with transfer at airport $j$ and back, is represented by the trigrams\\
\begin{minipage}{\textwidth}
\begin{verbatim}

  i i j 1
  i j k 1
  j k j 1
  k j i 1 

\end{verbatim}
\end{minipage}
with the first trigram \verb+i i j 1+ included only for calculating the teleportation weight in the PageRank analysis.

\subsection*{Journals} 
\label{sec:journal_data}

The journal citation data were extracted from JSTOR (www.jstor.org), a not-for-profit digital library that includes 2,227 journals and 8,227,537 citations among these journals. In order to capture memory, we extracted the underlying article-level citations.  This included 1,787,351 unique articles that cite at least one other JSTOR article or received a citation from another JSTOR article.   

JSTOR does not represent the full universe of scholarly content.  For example, the journal \textit{Nature} is not included in this subset.  In addition, physics, engineering, and computer science are not well represented in this corpus of articles.  However, it did offer several advantages.  First, JSTOR made its data available for research.  Second, the JSTOR corpus has both article- and journal-level data, which were necessary for building memory networks.   

For each article $A$ in JSTOR, we searched all articles $A_{\mathrm{out}}$ cited by $A$ and all articles $A_{\mathrm{in}}$ citing $A$. If we found at least one cited and one citing article, then, for each cited article, we picked a random citing article and formed the trigram
\begin{align}
A_{\mathrm{in}} \to A \to A_{\mathrm{out}},
\end{align}
which we mapped to the trigram between the corresponding journals
\begin{align}
J_{\mathrm{in}} \to J \to J_{\mathrm{out}}.
\end{align}
Finally, we aggregated all journal trigrams into the weighted memory network. By sampling memory networks many times, we found that choosing a random citing article did not affect our ranking or community detection results \cite{SI-bohlin2014ranking} (see Sec.\ 2). 

\subsection*{Patients} 
\label{sec:hospital_walks}

The patient data derive from a database of inpatient care at hospitals in Stockholm, Sweden, during 2001 and 2002 \cite{SI-liljeros2007contact}. The full dataset consists of 295,108 individuals who entered at least one of 702 wards at hospitals or nursing homes in 52 different locations. Our anonymised data were compiled by Fredrik Liljeros and are a sample of 365 days from the original data. Since we are only interested in patients who entered three or more wards, the data contain records from fewer individuals than the original data and are limited to patient movements between 402 wards. 

\subsection*{Taxis} 
\label{sec:taxi_data}

We included the taxi data because we were interested in analysing a real-world system lacking strong return flow.
With this data, we can contrast the dynamics in the other networks. 
The data are from GPS receivers in smartphones of taxi drivers from the Uber taxi
company in the San Francisco region \cite{SI-uber}. We further processed the data in the following way. First, because most of
the taxi traffic is in the metropolitan area of San Francisco, we limited
our analysis to the rectangle limited by longitudes 122.456 122.388 West and
latitudes 37.748 37.808 North. In that rectangle, we superimposed a
hexagonal grid of 20 x 20 hexagons, depicted in Supplementary Fig.\hs\ref{taxis_schematic}. The distance between opposite sides in each
hexagon of the grid is approximately 375 meters, or about the length of two city blocks. Indeed, we observed some
shorter trips in the dataset, but not many.
For each of 25,000 taxi trajectories in the data set, we built chains of 2-d integer coordinates corresponding to the hexagon where the taxi
is driving at a given moment. That is, geographical hexagons are the nodes of the
network, and there is a link between two hexagons if they share an edge.
Finally, we took triplets of consecutive hexagons visited by a taxi and
summarized them in trigrams, as described above. For calculating the weight of
each trigram, we used the number of times that a taxi trajectory contained the
trigram.

\begin{figure}[tbp]
    \centering
    \includegraphics[width=\columnwidth]{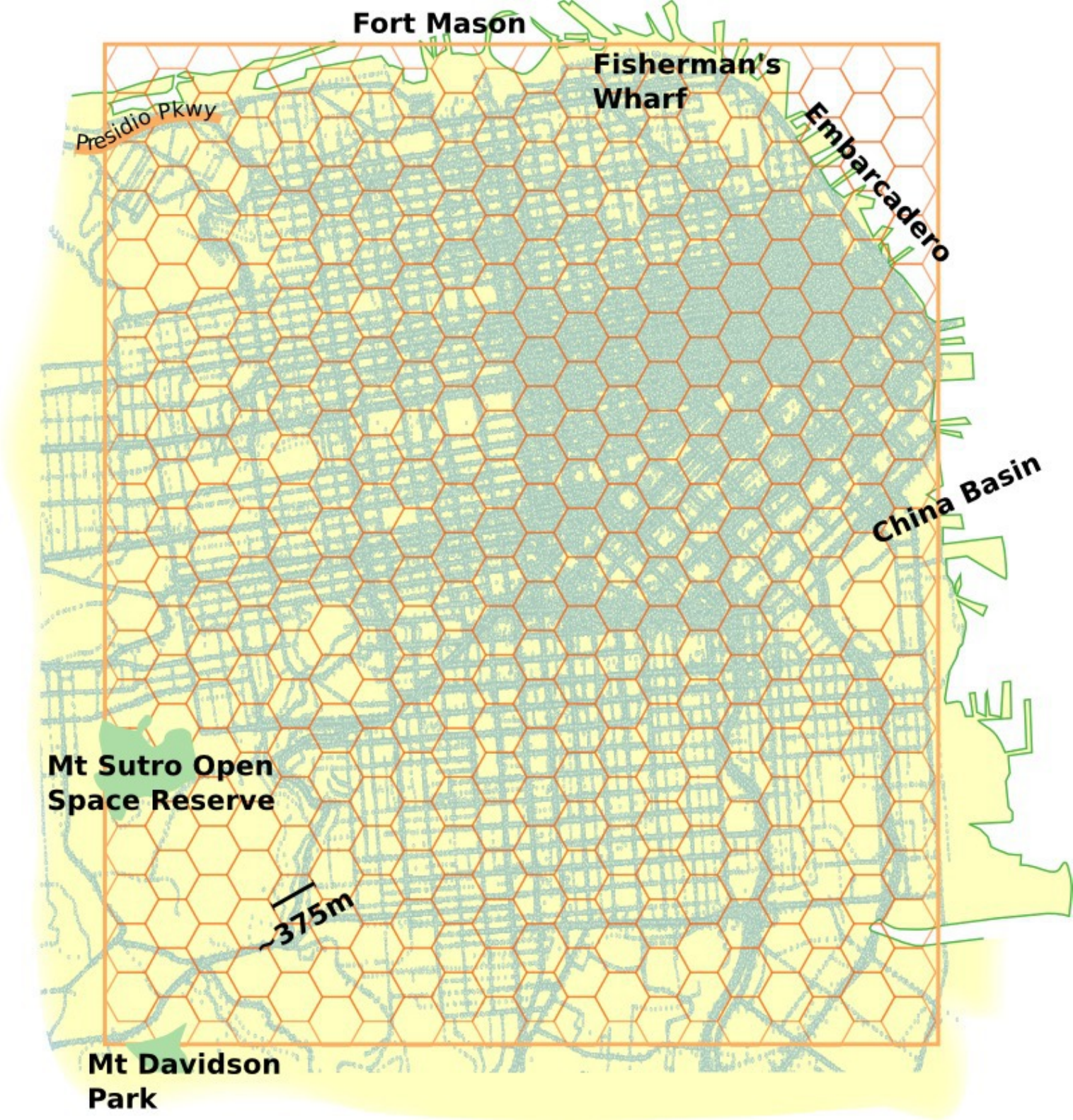}\\
    \caption{\textbf{Taxi traffic in the San Francisco urban centre.} We used the superimposed hexagonal grid to infer trigrams.    \label{taxis_schematic}}
\end{figure}

To validate that our results are not affected by the particular choice of grid structure, we also derived memory networks from 10 x 10 and 40 x 40 hexagonal grids. The results were qualitatively the same, with small quantitative differences. 

\subsection*{Emails} 
\label{sec:enron_emails}

The Enron email dataset \cite{SI-enron} comprises
emails between 146 users disclosed during the trial following the Enron scandal.
We aggregated all email addresses for each single user, and used a total of 116,525 messages between users as
links.  In order to establish succession between messages, we looked to
common  word-triplets present in the subject, overlap between the \emph{to} and \emph{from}
fields, and message date-time. Because we linked messages using their common
word-triplets, many messages were linked to many other messages through their subject line headings. We removed the
redundancy by leaving only one incoming link for each message, namely, the one
with the latest possible time-stamp. Finally, we broke the pathways into trigrams
as described above.

\section{Significance analysis with resampling}
\label{sec:bootstrapping}

To verify that our results are based on sufficient data, we performed bootstrap resampling of pathways for all summary statistics reported in the main manuscript and below, and surrogate data testing of the entropy rate to estimate the Markov order.

\subsection*{Bootstrap resampling}

For each dataset, we generated 100 bootstrap replicas by resampling the pathways with replacement and then constructed the networks from the replicas. Each dataset contains a large number of pathways distributed over a smaller number of unique pathways, such that each unique pathway has a weight given by the number of that unique pathway in the dataset. Therefore, we generated a single bootstrap memory network by first resampling the weights of all unique pathways from a multinomial distribution and then breaking the long pathways into trigrams (for patients, taxis, and emails, we only had access to trigrams and directly resampled their weights). For a given dataset, we used a multinomial distribution with as many categories as unique pathways in the dataset and with probabilities for the categories proportional to the weights of the unique pathways. From this multinomial distribution, we performed as many trials as the total number of pathways in the dataset and aggregated the outcome to resampled weights for all unique pathways.

For the journal network, we constructed the bootstrap replicas differently, because the journal memory network is not constructed from pathways but from chaining article citations.  The procedure described in Data acquisition and processing above involves a random step, and we simply generate the bootstrap replicas by repeating this procedure. Note that this procedure does not generate any variation in the standard network, but it is anyway the significance of the second-order Markov results that we are interested in.

For each set of summary statistics, we calculated the bootstrap confidence interval by ordering 100 bootstrap estimates and eliminated the ten smallest and ten largest estimates. In this way, the remaining estimates span the 90\% bootstrap percentile confidence interval. In general, we report the lower and upper limits of this interval. 
If there are many trigrams with only a few observations in the data, a summary statistic of the raw data can lie outside of the bootstrap confidence interval. Nevertheless, there can still be a memory effect. For a significant memory effect, what really matters is non-overlapping confidence intervals for M1 and M2 dynamics.

\subsection*{Surrogate data testing}

For the memory effects of individual nodes reported in Fig.\hs{2} of the main manuscript, we also performed hypothesis testing to verify that our results are based on sufficient data.
Our null hypothesis was that the flow is a first-order Markov process, and we used the conditional entropy at each node as a test statistic. Assuming that the null hypothesis is true, we estimated the probability that the conditional entropy of the second-order Markov process is at least as low as the observed value. We estimated this probability, the p-value, with bootstrap resampling \cite{SI-efron1994introduction} and rejected the null hypothesis if the p-value was lower than 0.10.

To resample the data at each node $i$, we used the so called symbolic surrogate procedure with constrained probabilities surrogates \cite{SI-van1998testing}. We first collected all trigrams $x_{1}x_{2}x_{3}$ that pass through node $i$ in the first step ($x_2=i$). These trigrams form a set of $n$ pairs
\begin{align}
\left\{(x_{1}^1,x_{3}^1),(x_{1}^2,x_{3}^2),\ldots, (x_{1}^n,x_{3}^n)\right\},
\end{align}
with all steps before and after node $i$. To resample these pairs, we randomized the order of the set of nodes visited before node $i$, $\left\{x_{1}^1,x_{1}^2,\ldots,x_{1}^n \right\}$ and created random pairings with the set of nodes visited after $i$, $\left\{x_{3}^1,x_{3}^2,\ldots,x_{3}^n \right\}$, thereby destroying any memory effect. After aggregating the transition probabilities for the random pairings, we calculated the conditional entropy for the randomized data. We repeated this procedure, random resampling and calculation of conditional entropy, as many times as needed to conclude that the p-value is lower or higher than 0.10. We used the Clopper-Pearson method \cite{SI-clopper1934use} with a 90\% confidence interval of the p-value to determine the stop condition.

For the air traffic data, we have sufficiently many long pathways to perform this analysis also for higher-order Markov models. For order $n$, we extracted all $n+1$-grams from the data. Since we use the conditional entropy as a test statistics, we estimate the average amount of information necessary to determine the destination of a passenger, given information about the sequence of airports the passenger has visited. In this way, the Markov order sets the horizon of how much information an observer at an airport has about passengers to determine their next step. For each Markov order $n$, we performed two tests: one in which we only included $n$-grams of length $n+1$, and one in which we also included all shorter $n$-grams of length 2, 3, \ldots $n$ from each pathway.  Excluding shorter $n$-grams corresponds to consider only passengers that already have visited at least $n$ airports, which we refer to as the \emph{maximum memory} of passengers at airports. Including shorter $n$-grams corresponds to consider all passengers and all their airport visits, which we refer to as \emph{typical memory} of passengers at airports.

For each Markov order $n$, we performed the same resampling procedure as described above. We split all $n+1$-grams into two sets, one that consists of the $n$ first airports of each $n+1$-gram and one that consists of the $n$ last airports of each $n+1$-gram. Then we generated resampled $n+1$-grams by randomly recombining the two sets such that each resampled $n+1$-gram begins with an $n$-gram from the first set and ends with an $n$-gram from the other set, with $n-1$ airports overlapping in the middle. In this way, the resampled $n+1$-grams will be of Markov order $n-1$. In the typical memory approach, we only resampled the longest $n+1$-grams, since all other 2-, 3-, \ldots $n$-grams are of order $n-1$ or lower. We repeated this resampling 100 times for each Markov order $n$ and compared the actual conditional entropies with the ones given by the null hypothesis that they are generated from Markov order $n-1$. Unlike in the main text where we weight the conditional entropies by PageRank, here we use the actual visit frequencies. 

\begin{figure}[ht]
    \centering
    \includegraphics[width=\columnwidth]{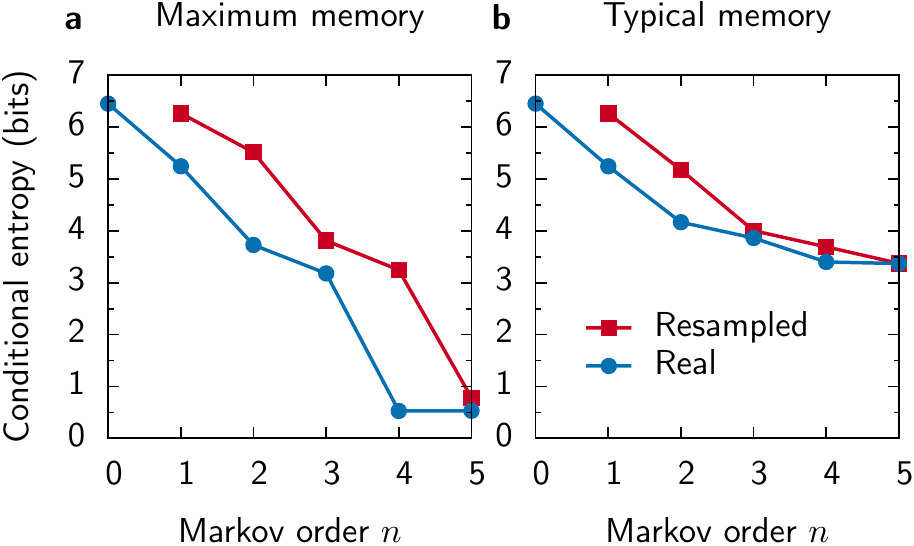}\\
    \caption{\textbf{Predicting the next destination in air traffic.} The entropy measures the uncertainty about a passengers next destination conditional on the sequence of already visited airports. The Markov order sets the maximum length of this memory. (\textbf{a}) Here we only consider passengers that have already visited at least $n$ airports. (\textbf{b}) Here we consider all passengers and all their airport visits. Resampled variance is less than the size of the symbols. \label{markovorder}}
\end{figure}

Supplementary Figure\hs\ref{markovorder} shows the results. Even if air traffic has statistically significant memory effects up to Markov order four (Supplementary Fig.\hs\ref{markovorder}a), shorter itineraries dominate (73\% of all itineraries are of length three or shorter) and a second-order Markov model accurately captures the typical memory dynamics (Supplementary Fig.\hs\ref{markovorder}b). For typical memory, the conditional entropy drops by 1.1 bits from first to second Markov order but only by 0.3 bits from second to third Markov order. Both results are statistically significant and we conclude that a second-order Markov model seems to successfully balance model complexity and accuracy. A more thorough model selection analysis is beyond the scope of this work. 

\begin{table*}[ht]
\caption{\textbf{Second-order Markov effects on constraints on flow} \label{constraintstable}}\vspace{-1em}
\flushleft
\setlength{\tabcolsep}{2.9pt}
{\small\begin{tabular}{@{}rllllllllllll@{}}\mytoprule\noalign{\smallskip}
{ Network} &\rule{0.5em}{0em}&  \multicolumn{2}{l}{ Two-step return (\%)} &\rule{0.5em}{0em}& \multicolumn{2}{l}{ Three-step return (\%)} &\rule{0.5em}{0em}& \multicolumn{2}{l}{ Entropy rate (bits)}\\ 
 && { M1} & { M2} && { M1} & { M2} && { M1} & { M2}\\ \noalign{\smallskip}
\cline{0-0}\cline{3-4}\cline{6-7}\cline{9-10}
\noalign{\smallskip}
Airports && 5.7 (5.7--5.7) & 47 (47--47) && 2.1 (2.1--2.1) & 0.63 (0.63--0.64) && 5.2 (5.2--5.2) & 3.4 (3.4--3.4) \\ 
Cities && 6.5 (6.5--6.5) & 48 (48--48) && 2.8 (2.8--2.8) & 0.62 (0.62--0.62) && 4.7 (4.6--4.7) & 3.5 (3.5--3.5) \\
Journals && 11 (11--11) & 21 (21--21) && 4.7 (4.7--4.7) & 5.4 (5.4--5.5) && 4.5 (4.5--4.5) & 3.5 (3.5--3.5) \\
\noalign{\smallskip}
Patients && 16 (16--18) & 54 (51--55) && 1.9 (1.7--2.1) & 3.4 (2.0--3.2) && 3.0 (2.5--2.6) & 1.0 (0.92--1.0) \\
Taxis && 20 (20--20) & 10 (10--11) && 6.8 (6.8--6.9) & 10 (10--10) && 2.2 (2.2--2.2) & 1.1 (1.1--1.1) \\
Emails && 14 (14--15) & 58 (55--58) && 5.2 (5.1--5.5) & 2.7 (2.2--3.2) && 3.0 (2.8--2.9) & 1.3 (1.1--1.2) \\
\mybottomrule
\end{tabular}
\parbox{0.75\textwidth}{\footnotesize\raggedright M1 and M2 for results obtained with a first- and a second-order Markov model, respectively. Values in parentheses represent the 10th and 90th percentiles from the bootstrap analysis. All node averages are for physical nodes.}
}
\end{table*}

Finally, in Supplementary Table\hs\ref{constraintstable} we report the conditional entropies of first- and second-order Markov dynamics together with the two-step and three-step return rates. Here we also complement the results reported in Table\hs{1} of the main manuscript with the 10th and 90th percentiles of the bootstrap values. The bootstrapping shows that the memory effects are significant. For airports and cities, the data are sufficiently rich that more than two digits are significant. For the patient data, however, many pathways through the hospitals are used by only one or a few patients. Therefore, the bootstrap estimates vary more and sometimes exclude the summary statistics of the raw data. Nevertheless, there is a significant effect of second-order Markov dynamics in all cases.

\section{Community detection of memory networks}

We have seen that a second-order Markov model has important effects on dynamic processes on networks. To better understand these effects, we simplified the dynamics and highlighted the important structures of the dynamics with community detection, currently the best way to comprehend dynamics on a large scale \cite{SI-fortunato2010community}. With community detection, we can compare the structure of first- and second-order Markov dynamic. Since we are interested in the dynamics, we have chosen to work with the flow-based map equation framework \cite{SI-rosvall2008maps}. Alternative flow-based methods exist \cite{SI-simonsen2004diffusion,SI-delvenne2010stability}, but the map equation framework easily allows us to maintain the mechanics of the method and only modify the dynamics. That is, we can cluster physical nodes and use memory nodes for controlling the dynamics. This advantageous feature allows us to efficiently compare the first- and second-order Markov dynamics. Since we are interested in overlapping modules, we build our new method on a generalization of the map equation to overlapping modules\cite{SI-esquivel2011compression}.

\subsection*{The map equation}

The map equation framework is an information-theoretic approach that takes advantage of the duality between compressing data and finding regularities in the data. Given module assignments $\mathsf{M}$ of all nodes in the network, the map equation measures the description length $L(\mathsf{M})$ of a random walker who moves within and between modules from node to node by following the links between the nodes \cite{SI-rosvall2009map}: 
\begin{align}\label{map}
L(\mathsf{M}) = q_{\curvearrowleft} H(\mathcal{Q}) + \sum_{i=1}^{m}p_{\circlearrowright}^iH(\mathcal{P}^i)
\end{align}
Here the entropy $H(\mathcal{Q})$ measures the average per-step description length of movements between modules derived from module-enter frequencies $\mathcal{Q}$ of all $m$ modules and $H(\mathcal{P}^i)$ measures the average per-step description length of movements within module $i$ derived from node-visit and module-exit frequencies $\mathcal{P}^i$. The description lengths are weighted by their frequency of use, $q_{i \curvearrowleft}$ and $p_{\circlearrowright}^i$, respectively. The visit frequencies can be obtained by first calculating the PageRank of nodes and links with smart teleportation as described in the Methods section of the main text, or directly from the data if the links represent flow themselves.

In any case, finding the optimal partition of the network by assigning each node to one or more modules corresponds to testing different node assignments and picking the one that minimizes the map equation.

The challenge is to handle the large search space of possible solutions when nodes can be assigned to any number of overlapping modules. Therefore, we limit the search space here and only allow each memory node to be assigned to a single module. In this way, the efficient \cite{SI-lancichinetti2009community} search algorithm for hard partitions in \emph{Infomap} \cite{SI-rosvall2010mapping,SI-infomap} can be used with only small modifications. Instead of applying the search algorithm on the standard network, we apply it on the memory network that contains transition information between memory nodes (links). The method is thus a form of link partitioning \cite{SI-evans2009line,SI-ahn2010link}. The search algorithm initiates each memory node in its own module and proceeds as \emph{Infomap} for hard partitions of regular nodes\cite{SI-rosvall2010mapping}, with one important difference: When two or more memory nodes of the same physical node are assigned to the same module, the description length must capture the fact that the memory nodes share the same codeword. That is, to obtain the visit frequency of a physical node in a module, we sum the visit frequencies of all memory nodes of that physical node in the module. We then use this visit frequency to derive the optimal codeword length. This procedure is essential to ensure that the map equation measures the optimal description length of a random walker navigating between physical nodes. In this way, the compression algorithm remains the same and only the dynamics change. Moreover, we ensure that the community detection results only depend on memory effects by representing first-order Markov dynamics flow in a memory network, with each memory node having the out-links of its corresponding physical node in the standard network. 

\begin{figure*}[ht]
\centering
\includegraphics[width=\textwidth]{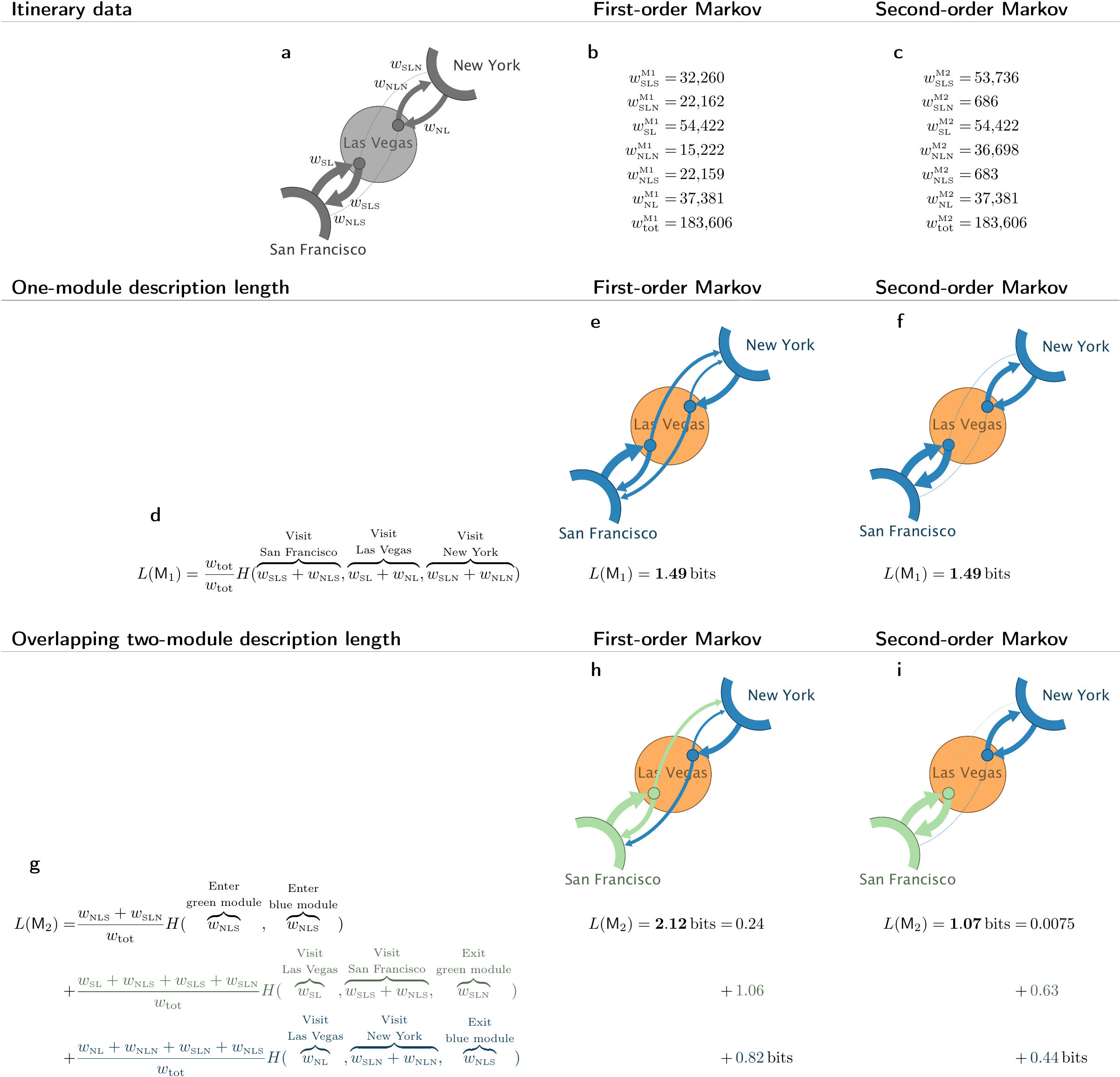}\\
\caption{\label{lasvegasclustering} \textbf{Second-order Markov dynamics reveal overlapping modules.} Itineraries weighted by passenger number from the data presented in Sec.\ 1, restricted to transfer traffic and trigrams that include San Francisco, Las Vegas, and New York. (\textbf{a}) Legend with labelled links between memory nodes. For example, $w_{\text{\tiny NLN}}$ represents pathways that start in New York, continue to Las Vegas, and return to New York. (\textbf{b}) First-order Markov dynamics represented with links between memory nodes. (\textbf{c}) Second-order Markov dynamics represented with links between memory nodes. (\textbf{d}) The map equation for the one-module solution $\textsf{M}_1$ in \textbf{e}-\textbf{f}. Normalization of the weights in the entropy is implicit. (\textbf{e}) The description length of the one-module solution $\textsf{M}_1$ with memoryless flow. (\textbf{f}) The description length the a one-module solution $\textsf{M}_1$ with second-order Markov dynamics. (\textbf{g}) The map equation for the overlapping two-module solution $\textsf{M}_2$ in \textbf{h}-\textbf{i}. First line gives the average description length of movements between the two modules. Second line gives the average description length of movements within the green module. Third line gives the average description length of movements within the blue module. (\textbf{h}) The description length of the overlapping two-module solution $\textsf{M}_2$ with first-order Markov dynamics. (\textbf{i}) The description length of the overlapping two-module solution $\textsf{M}_2$ with second-order Markov dynamics.}
\end{figure*}

Supplementary Figure\hs\ref{lasvegasclustering} illustrates the mechanics of the map equation for first- and second-order Markov dynamics. The first-order passenger trigrams in b are derived from the actual trigrams in c in two steps. We first derived the normalized out-links of the two memory nodes in Las Vegas. Since these memory nodes should represent first-order Markov dynamics, their out-links are identical and equal to the proportion of passengers flying to San Francisco and New York, respectively. We then multiplied these transition probabilities by the number of passengers that arrive in Las Vegas from San Francisco and New York, respectively. In this way, for example,
\begin{align}
w_{\text{\tiny SLS}}^{{\text{\tiny M1}}} = w_{\text{\tiny SL}}^{{\text{\tiny M2}}}\frac{w_{\text{\tiny SLS}}^{{\text{\tiny M2}}} + w_{\text{\tiny NLS}}^{{\text{\tiny M2}}}}{w_{\text{\tiny SL}}^{{\text{\tiny M2}}} + w_{\text{\tiny NL}}^{{\text{\tiny M2}}}},
\end{align}
where
\begin{align}
w_{\text{\tiny SL}}^{{\text{\tiny M2}}} = w_{\text{\tiny SLS}}^{{\text{\tiny M2}}} + w_{\text{\tiny SLN}}^{{\text{\tiny M2}}}
\end{align}
and
\begin{align}
w_{\text{\tiny NL}}^{{\text{\tiny M2}}} = w_{\text{\tiny NLN}}^{{\text{\tiny M2}}} + w_{\text{\tiny NLS}}^{{\text{\tiny M2}}}.
\end{align}
Further, 
\begin{align}
w_{\mathrm{tot}}^{{\text{\tiny M1}}} = w_{\mathrm{tot}}^{{\text{\tiny M2}}} = 2w_{\text{\tiny SL}}^{{\text{\tiny M2}}} + 2w_{\text{\tiny NL}}^{{\text{\tiny M2}}}
\end{align}
corresponds to the total passenger weight involved in the set of two consecutive flight legs illustrated in Supplementary Fig.\hs\ref{lasvegasclustering}.

\begin{table}[ht]
\caption{\textbf{Memory reveals multidisciplinary journals in the scholarly literature}\label{table:science}}\vspace{-1em}
\flushleft
\setlength{\tabcolsep}{2.9pt}
\begin{small}
\begin{tabular}{@{}rllllllllllll@{}}\mytoprule\noalign{\smallskip}
{Field} && \multicolumn{2}{l}{PNAS} &\rule{0.5em}{0em}& \multicolumn{2}{l}{Science} &\rule{0.5em}{0em}& \multicolumn{2}{l}{Ecology} &\rule{0.5em}{0em}& \multicolumn{2}{l}{Plant Cell}\\ \noalign{\smallskip}
&& {M1} & {M2} && {M1} & {M2} && {M1} & {M2} && {M1} & {M2} \\ \noalign{\smallskip}
\cline{0-0}\cline{3-4}\cline{6-7}\cline{9-10}\cline{12-13}\noalign{\smallskip}
Ecology && $-$ & 13
&& $-$ & 29
&& 100 & 100
&& $-$ & $-$\\
Cell biology && 100 & 80
&& 100 & 68
&& $-$ & $-$
&& 100 & 100 \\
Mathematics && $-$ & 4.6
&& $-$ & $-$
&& $-$ & $-$
&& $-$ & $-$ \\
\noalign{\smallskip}
Statistics && $-$ & 1.5
&& $-$ & $-$
&& $-$ & $-$
&& $-$ & $-$ \\
Anthropology && $-$ & $-$
&& $-$ & 1.6 
&& $-$ & $-$
&& $-$ & $-$\\
Others && $-$ & 0.38$^*$
&& $-$ & 1.4$^\dagger$
&& $-$ & $-$
&& $-$ & $-$\\
\mybottomrule
\end{tabular}\\
\parbox{0.47\textwidth}{\footnotesize\raggedright The relative assignment to each field in percentage. Number of other fields a journal is assigned to in parenthesis. $^*$In 1 other field. $^\dagger$In 7 other fields.}
\end{small}
\end{table}

\subsection*{Memory and heuristic algorithms}

The link clustering \cite{SI-ahn2010link} and clique percolation \cite{SI-palla2005uncovering} methods can be seen as trying to account for second-order Markov dynamics. They operate by increasing connectivity within modules and decreasing connectivity between modules, albeit in different ways. Below we establish this flow interpretation of the two methods. 

\begin{figure}[ht]
\centering
\includegraphics[width=\columnwidth]{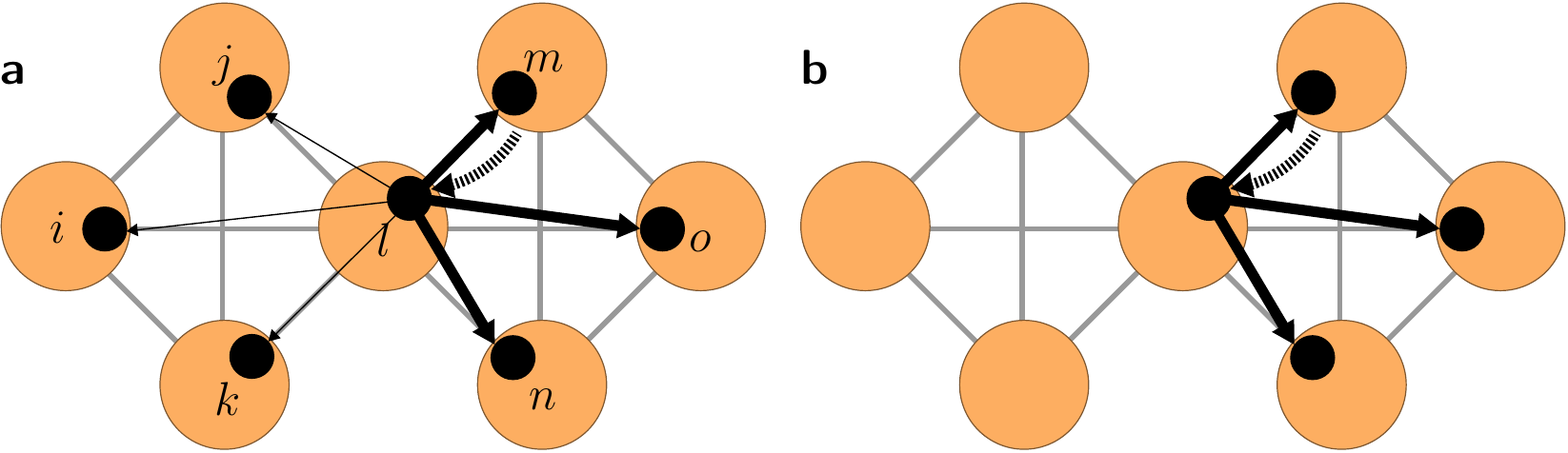}\\
\caption{\label{cliquepercolation} \textbf{Interpreting link clustering and clique percolation as flow-based methods with second-order Markov dynamics.} (\textbf{a}) Link clustering interpreted as a flow-based method with second-order Markov dynamics: The similarity between link $m \to l$ and other links connected to node $l$ represented as similarity weighted out-links from memory node $\protect\vv{ml}$ to connected memory nodes. (\textbf{b}) Clique percolation interpreted as a flow-based method with second-order Markov dynamics: Possible link sides of adjacent triangles sharing the link $m \to l$ represented as unweighted out-links from memory node $\protect\vv{ml}$.}
\end{figure}

The link clustering method measures the similarity between two links $k \to i$ and $k \to j$ connected at a common ``keystone'' node $k$ as the ratio between all shared nodes and the total number of nodes reached in two steps from the keystone node via the links. That is, the similarity is $|n_+(i) \cap n_+(j)|/|n_+(i) \cup n_+(j)|$, where $n_+(i)$ is the set of neighbours of $i$. Supplementary Figure\hs\ref{cliquepercolation}a illustrates the similarities between nodes. With the keystone node in the middle, the dashed link is connected to five other links. The similarity is 1 with the links in the same module, because they share all nodes two steps from the keystone node via the links. However, the similarity is only 1/7 with the links in the other module, because they only share the keystone node of all seven nodes that can be reached in two steps from the keystone node via the links. With these link weights between links, it is clear that the link clustering method identifies two modules overlapping at the node in the centre. How can we interpret this machinery as constraints on flow? With similarities represented as links between memory nodes, as in Supplementary Fig.\hs\ref{cliquepercolation}a, we see that a random walker would rarely switch between the modules. Conditional on being at the node in the centre, and assuming that the link self-similarity is 1, the transition rate decreases from 1/2 to 1/7 with weights derived according to the link community procedure. In this way, the persistence time increases in the two modules and allow for efficient compression of the flow.

Clique percolation identifies a module as the maximum set of nodes that can participate in a percolation of adjacent cliques. A clique is a fully connected sub-graph and two cliques are adjacent if they share all nodes but one. Here we consider sub-graphs of size three, triangles, such that two triangles are adjacent if they share two nodes or, equivalently, one side. Supplementary Figure\hs\ref{cliquepercolation}b illustrates. Since a triangle can percolate between the leftmost four nodes or the rightmost nodes, the method identifies two modules overlapping at the node in the centre. How can we interpret this process as constraints on flow? Assume that a random walker steps from node $m$ to node $l$. The two adjacent triangles that share this link $m \to l$ have link sides $l \to o$ and $l \to n$, respectively, connected to node $l$. We now restrict the random walker to only move along those links, or back from where it came. If we use memory nodes to represent these constraints, as in Supplementary Fig.\hs\ref{cliquepercolation}b, the transition rate between the two modules drops to zero. In fact, the random walker will be blocked by the very same module boundaries as given by the clique percolation method. Again, the persistence time increases in the two modules and allow for efficient compression of the flow.

By interpreting the machinery of the two methods as constraints on flow, we see that they can be seen as trying to infer second-order Markov dynamics from the structure of the standard network. Moreover, those constraints give longer persistence time in modules. And indeed, as we have seen from using real data of second-order dynamics, the persistence time in modules does increase when accounting for higher-order memory effects. As a result, this flow interpretation establishes an interesting connection between two heuristic methods that operate on standard networks and our inherently flow-based method that operates on memory networks. We conclude that this flow interpretation provides more principled grounds for link clustering and clique percolation.

\subsection*{Results of the statistical analysis}

\begin{table*}[ht]
\caption{\textbf{Second-order Markov effects on community detection and ranking} \label{clusteringtable}}\vspace{-1em}
\flushleft
\setlength{\tabcolsep}{2.9pt}
{\small\begin{tabular}{@{}rllllllllllllll@{}}\mytoprule\noalign{\smallskip}
{ Network} && \multicolumn{2}{l}{ Module size (\%)} &\rule{0.5em}{0em}&  \multicolumn{2}{l}{ Module assignments}&\rule{0.5em}{0em}&  \multicolumn{1}{l}{ Compression gain (\%)} && \multicolumn{1}{l}{ Ranking difference (\%)} \\ 
 && { M1} & { M2} && { M1} & { M2} && { M1\raisebox{0.1ex}{$\to$}M2} && { M1\raisebox{0.1ex}{$\to$}M2}\\ \noalign{\smallskip}
\cline{0-0}\cline{3-4}\cline{6-7}\cline{9-9}\cline{11-11}
\noalign{\smallskip}
Airports && 93 (93--93) & 5.1 (5.0--5.1) && 1.2 (1.2--1.2) & 6.2 (6.2--6.3) && 13 (9.7--9.8) && 8.2 (8.1--8.2) \\ 
Cities && 32 (32--32) & 5.3 (5.4--7.7) && 1.8 (1.8--1.8) & 3.7 (3.6--3.7) && 4.7 (4.3--4.3) && 3.7 (3.7--3.7) \\
Journals && 14 (14--14) & 15 (15--15) && 1.8 (1.8--1.8) & 3.4 (3.3--3.4) && 4.7 (10--11) && 9.7 (9.5--9.8) \\
\noalign{\smallskip}
Patients && 7.3 (5.2--6.9) & 1.9 (1.7--2.1) && 5.0 (4.4--4.7) & 4.7 (4.4--5.0) && 30 (23--28) && 22 (24--28)\\
Taxis && 3.1 (2.9--3.3) & 2.2 (2.2--2.4) && 1.5 (1.5--1.6) & 1.7 (1.7--1.8) && 6.5 (6.5--7.2) && 6.5 (6.5--7.1)\\
Emails && 12 (11--13) & 5.8 (4.7--6.3) && 1.3 (1.4--1.6) & 3.0 (2.5--2.7) && 26 (23--27) && 18 (20--24)\\
\mybottomrule
\end{tabular}
\parbox{0.85\textwidth}{\footnotesize\raggedright M1 and M2 for results obtained with first- and second-order Markov models, respectively. Values in parentheses represent the 10th and 90th percentiles from the bootstrap analysis. All node averages are for physical nodes.}
}
\end{table*}

We summarize with the results of the community detection analysis and the closely related ranking in Supplementary Table\hs\ref{clusteringtable}. In addition to the results already presented in Table\hs{1} of the main manuscript, here we also report the 10th and 90th percentiles from the bootstrap analysis. Because of the greedy and stochastic nature of the search algorithm, the confidence intervals are wider than for the entropies and return rates reported in Supplementary Table\hs\ref{constraintstable}. Nevertheless, there is a significant difference between the structure of first- and second-order Markov dynamics. In a second-order Markov model, the dynamics are confined in smaller and more overlapping modules.

\section{Modelling second-order Markov effects} \label{sec:Model}

So far, we have used empirical data and studied the effects of second-order Markov dynamics on community detection, ranking and spreading processes. Here we outline a different line of research aimed at identifying simple mechanisms for explaining the effects of memory in networks. The modelling approach can be seen as an initial step in bringing together complementary knowledge from previously disconnected areas of research: Biologists have used empirical data to model animal movements in 2D with correlated random walks \cite{SI-kareiva1983analyzing,SI-bovet1988spatial,SI-bergman2000caribou}, computer scientists have used web logs to predict web surfer behaviour with higher-order Markov models \cite{SI-pirolli1999distributions,SI-chierichetti2012web}, and physicists have used theoretical models to study dynamics on networks with biased random walks \cite{SI-fronczak2009biased,SI-burda2009localization,SI-boldi2012arc}. By combining the empirical work for prediction with the theoretical work for mechanistic understanding, we are in a good position to better understand the effects of memory in integrated systems. 

To combine the approaches, we developed a simple network memory model that, fitted to data, can capture some basic features of second-order Markov dynamics in real systems. In addition to its explanatory power, the memory model also summarizes the dynamics of a system and makes it easy to compare the dynamics between different systems. Below, we first describe the memory model, then we show a procedure for fitting the model parameters to real data, and finally we illustrate, with ranking as an example, how this modelling approach can be used for a mechanistic understanding of the effects of second-order Markov dynamics. 

\subsection*{The memory model}

We build a tractable model for memory dynamics by coarse-graining the description of second-order Markov data. We define three different types of transition between nodes: a {\em return step} $r_2$, where the walker goes from $i$ to $j$ to $i$; a {\em triangular step} $r_3$, where the walker goes from $i$ to $j$ to a neighbour of $i$; and an {\em exploratory step} $r_{3<}$, for which the destination of the step is neither of those previously described. These events correspond to transitions of the types $\vv{ij} \rightarrow \vv{ji}$, $\vv{ij} \rightarrow \vv{j\sigma(i)}$, and $\vv{ij} \rightarrow \vv{j\gamma(i)}$, where $\sigma(i)$ is a member of the set of neighbours of $i$ and $\gamma(i)$ is a member of the set of nodes different from $i$ and the set of neighbours of $i$ (in principle, we can chose the set of either in-neighbours, out-neighbours, or neighbours in an undirected sense. In this discussion, we chose the latter option for the sake of simplicity). 

\begin{figure}[ht]
\centering
\includegraphics[width=0.5\columnwidth]{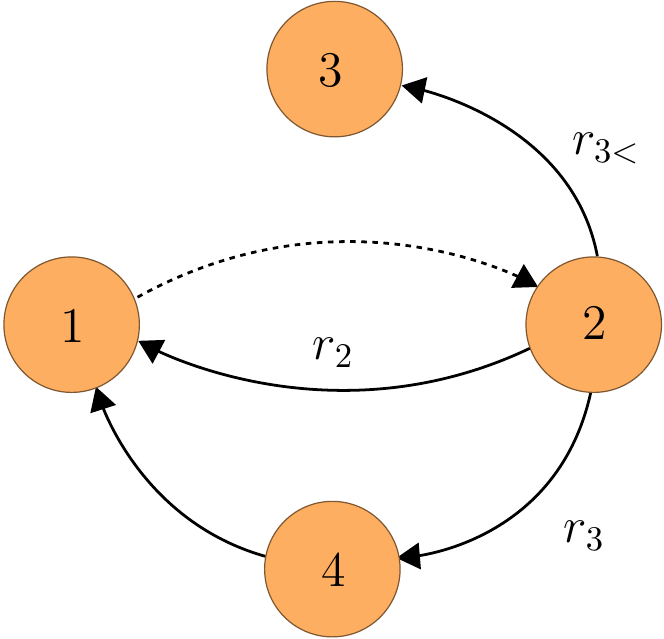}\\
\caption{\textbf{Illustration of the memory model.} After performing a jump along the dashed link, a walker can either perform a return step, with probability $r_2$, a triangular step, with probability $r_3$, or an exploratory step, with probability $r_{3<}$.\label{figabc}}
\end{figure}

The memory model is defined by assigning a different prevalence $w(\vv{ij} \rightarrow \vv{jx})$ to each type of transition, $r_2$, $r_3$, and $r_{3<}$, respectively (see Supplementary Fig.\hs\ref{figabc}), where the index underlines the length of the cycle associated with the process. For instance, the probability of performing a return step from $\vv{ij}$, and thus  of observing a cycle of length $2$, is
\begin{equation}
p_{\rm return} = r_2/(r_2 + r_3 |\sigma(i)| + r_{3<} |\gamma(i)|)
\end{equation}
if $\vv{ji}$ exists; otherwise, it is zero. $|\sigma(i)|$ is the number of elements in $\sigma(i)$. Tuning the values of $r_x$ gives more or less importance to each type of step. Importantly, the values of the parameters can easily be evaluated in empirical data, as shown below. Without loss of generality, we impose the constraint $r_2 + r_3 + r_{3<} =1$, such that $r_x$ can be understood as the probability of an event of type $x$ occurring. The memory model is described by transition probabilities $\hat{p}(\vv{ij} \rightarrow \vv{jk})$, approximating the original transition probabilities $p(\vv{ij} \rightarrow \vv{jk})$, and depending on the value of the above parameters and on the type of transitions between $\vv{ij}$ and  $\vv{jk}$. 

\subsection*{Fitting the memory model}
\label{fitting_subsection}

We now seek to find parameter values of $r_2$, $r_3$, and $r_{3<}$ to model dynamics as close as possible to observed M2 data. The model is thus fitted by minimizing the difference between the observed transition probabilities $p(\vv{ij} \rightarrow \vv{jk})$ measured from the trigrams and the transition matrix of the model $\hat{p}(\vv{ij} \rightarrow \vv{jk})$. To do so, we look for the values of $r_2$ and $r_3$, minimizing the Kullback$-$Leibler (KL) divergence
  \begin{equation}
 D_{\textrm{KL}}=   \sum_{\vv{ij}}   \pi(\vv{ij})    \sum_{\vv{jk}} p(\vv{ij} \rightarrow \vv{jk}) \log{ \frac{p(\vv{ij} \rightarrow \vv{jk})}{ \hat{p}(\vv{ij} \rightarrow \vv{jk})}},
 \end{equation}
 where $ \pi(\vv{ij}) $ is, as before, the PageRank of node $\vv{ij}$. Minimizing the KL divergence is equivalent to maximizing the log-likelihood \cite{SI-akaike1998information}, but since we use the conditional entropy to quantify the constraints on flow, we find it natural to use the information-theoretic KL divergence as the objective function.

In practice, in order to minimize the KL divergence, we first analytically derive its partial derivatives with respect to $r_2$ and $r_3$ ($r_{3<}$ does not appear in the expression of the KL divergence because of the normalization $r_{3<} = 1-r_2 -r_3)$. Then we set the derivatives to zero and iteratively solve the two coupled equations with a simple bisection method. 

\begin{figure}[ht]
\centering
\includegraphics[width=\columnwidth]{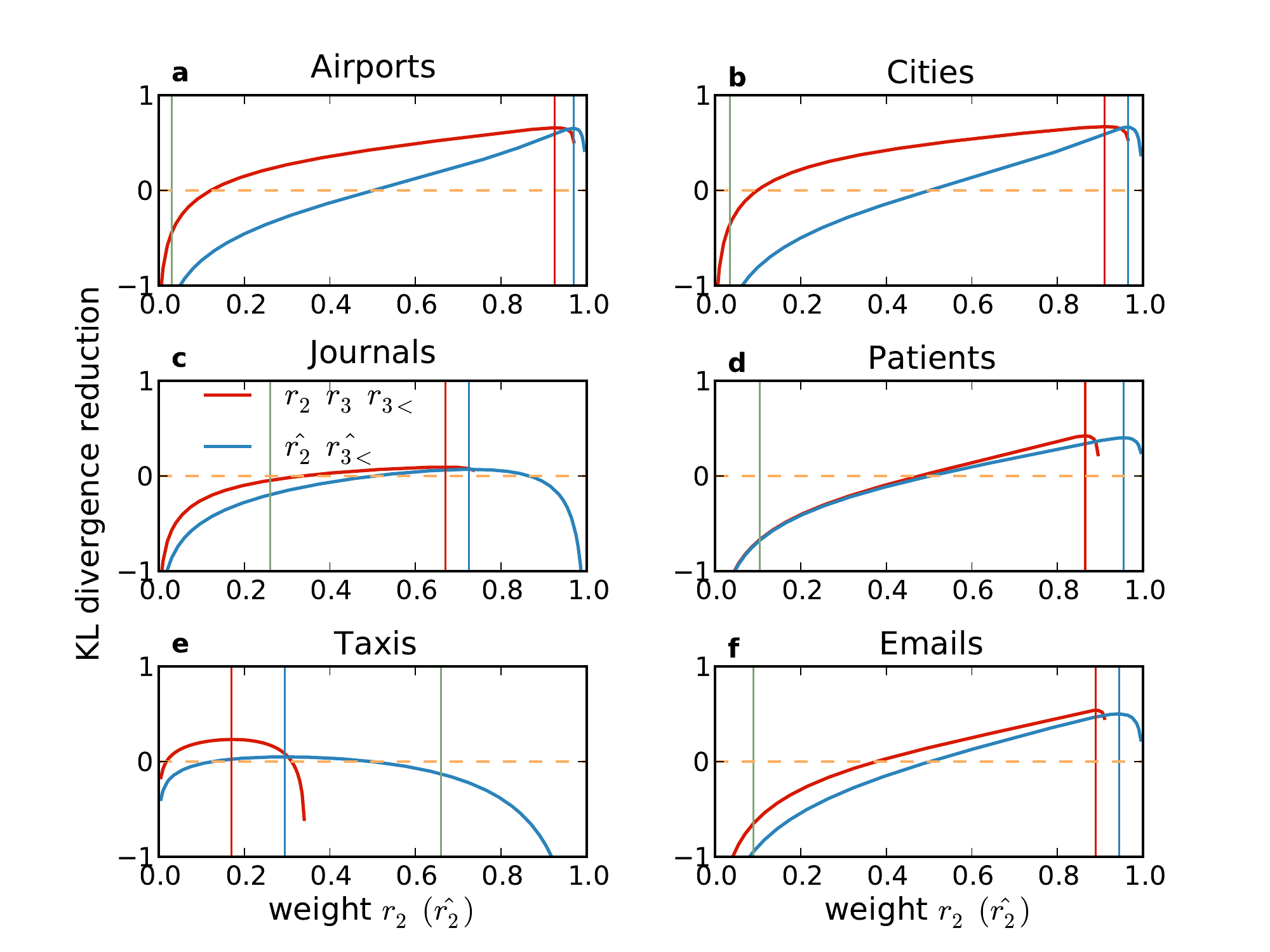}\\
\caption{\textbf{Fitting the memory model.} The plots show the  ${\textrm{KL}}$ divergence reduction in the memory model with all parameters $r_2$, $r_3$, $r_{3<}$ (red), and with restricted parameters $\hat{r_2}$, $\hat{r_{3<}}$ (blue) model. The vertical lines correspond to $r_2$ (red), $r_3$ (green) and $\hat{r_2}$ (blue).}
\label{kl_fig}
\end{figure}

The parameters found by performing this optimization are summarized in Supplementary Table\hs\ref{klmodel_table}. The table also reports values for $\hat{r_2}$, the best parameter in a simplified model where we impose that $r_3=r_{3<}$, and thus only differentiate between return steps and other type of steps. We quantified the relative ${\textrm{KL}}$ reduction as 1 minus the ratio between the optimized ${\textrm{KL}}$  divergence and the unbiased first-order Markov model with $r_2 + r_3 + r_{3<}=1/3$ ($\hat{\textrm{KL}}$ refers to the simplified model); high values of the ${\textrm{KL}}$ reduction imply a higher relative gain of fitting the data with the memory model over a first-order Markov model. In most cases, we observe that the inclusion of memory through the parameters $r_2$, $r_3$, and $r_{3<}$ significantly improves the accuracy of the modelling. 
A majority of networks show a high value of $r_2$, as expected. We also observe that the reduction induced by relaxing the constraint $r_3=r_{3<}$ tends to be small, suggesting that the definition of return steps is the most important ingredient in producing realistic pathways. Supplementary Figure\hs\ref{kl_fig} shows the ${\textrm{KL}}$ divergence reduction when tuning $r_2$ and $\hat{r_2}$.

\begin{table*}[ht]
\raggedleft\caption{\textbf{Fitted model parameters and relative reduction in KL divergence over a first-order Markov model} \label{klmodel_table}}\vspace{-1em}
\flushleft
{\small
\begin{tabular}{rllllllllllll}\mytoprule\noalign{\smallskip}
Network &\rule{1em}{0em}& $r_2 (\%)$ &\rule{1em}{0em}& $r_3 (\%)$ &\rule{1em}{0em}& $r_{3<} (\%)$ &\rule{1em}{0em}& ${\textrm{KL}} $ red.\ $(\%)$ &\rule{1em}{0em}& $\hat{r_2} (\%)$ &\rule{1em}{0em}& $\hat{\textrm{KL}} $ red.\ $(\%)$\\
\cline{0-0}\cline{3-7}\cline{9-9}\cline{11-11}\cline{13-13}
\noalign{\smallskip}
Airports && 93 (93-93) && 2.8 (2.8-2.8) && 4.5 (4.5-4.5) && 66 (66-66) && 97 (97-97) && 65 (65-65)\\
Cities && 91 (91-91) && 3.2 (3.2-3.2) && 5.6 (5.6-5.7) && 67 (67-67) && 96 (96-96) && 66 (66-66)\\
Journals && 67 (67-67) && 25 (25-25) && 7.6 (7.2-7.6) && 9.6 (9.4-9.6) && 72 (72-72) && 7.0 (6.9-7.0)\\
Patients && 86 (85-90) && 10 (7.4-11) && 3.4 (2.8-3.5) && 40 (40-45) && 95 (95-95) && 40 (40-45)\\
Taxis && 17 (16-17) && 66 (65-66) && 17 (17-18) && 23 (22-23) && 29 (29-30) && 4.8 (4.8-5.0)\\
Emails && 89 (88-89) && 8.9 (8.4-9.8) && 1.9 (1.8-2.2) && 54 (51-55) && 94 (94-95) && 50 (47-51)\\
\mybottomrule
\end{tabular}
\parbox{0.7\textwidth}{\footnotesize\raggedright Values in parentheses are the 10th and 90th bootstrap percentiles. $\hat{\textrm{KL}}$ refers to the simplified model, where we only account for returning steps.}
}
\end{table*}

It is worth mentioning that the optimization procedure provides a unique solution because the search landscape is indeed very smooth. For example, Supplementary Fig.\hs\ref{kl_fig} shows that the KL divergence has a unique minimum (the reduction has a maximum) as a function of $r_2$ with $r_3$ fixed. Also the reversed scenario, fixed $r_2$ and variation in $r_3$, has a similar smooth form with a unique extremum. Moreover, the bootstrap analysis shows that the optimal parameters are very robust. 

\subsection*{Analytical analysis of second-order Markov effects on ranking}

Here we use the model to illustrate the effects of second-order Markov dynamics on ranking. We use schematic networks and, for simplicity, we focus on unweighted networks. In this case, $W(i \to j)$ is simply the adjacency matrix $A_{ij}$ of the network, with $A_{ij}=1$ if there is a link going from $i$ to $j$ and zero otherwise. It is also useful to introduce the in- and out-degrees of each node defined by $\sigma_j^{\rm in}=\sum_i A_{ij}$ and $\sigma_i^{\rm out}=\sum_j A_{ij}$.
As a first step, we focus on the basic case when each type of transition is equally probable, $r_2 + r_3 + r_{3<}=1/3$. In this case, the memory model is equivalent to a standard Markov random walk on the physical network. To show this, we first note that (7) in the main manuscript reduces to 
\begin{equation}
\label{standard1}
P(\vv{jk};t+1) = \sum_{i} P(\vv{ij};t) \frac{A_{jk} }{\sigma_j^{\rm out}}.
\end{equation}
It is straightforward to show that the stationary solution of the process is $\pi(\vv{jk})=1/L$ if the graph is Eulerian ($\sigma_j^{\rm out}=\sigma_j^{\rm in}$ for all $j$). One thus recovers the well-known result that the stationary probability of finding a walker on a node is proportional to its degree in undirected networks. If the underlying network is strongly connected, this is the only stationary solution of the process. 
By using the fact that 
\begin{equation}
P(j;t)=\sum_{i} P(\vv{ij};t), 
\end{equation}
and summing over $j$ in Supplementary Equation (\ref{standard1}), we find that
\begin{equation}
\label{standardnode}
P(k;t+1) = \sum_{j} P(j;t) \frac{A_{jk} }{\sigma_j^{\rm out}},
\end{equation}
thus recovering the standard master equation for a random walk process, driven by the transition matrix $A_{jk} / \sigma_j^{\rm out}$.

If the standard random walk is ergodic on a graph, the memory model is also ergodic on the same graph for any value of $r_2$, $r_3$, and $r_{3<}$, as long as each parameter is strictly positive, such that no transition is forbidden by the bias.
In systems where ergodicity is not verified, we use link teleportation as described above. The robustness of link teleportation under variations of the teleportation probability $1-\alpha$ \cite{SI-lambiotte2012ranking} is clear, after noting that the stationary solution of the first-order Markov process  is $\pi(\vv{jk})=1/L$ when each node of the physical network has the same in-degree and out-degree (e.g., if the network is undirected), independently of $\alpha$.

\begin{figure}[ht]
\centering
\includegraphics[width=0.8\columnwidth]{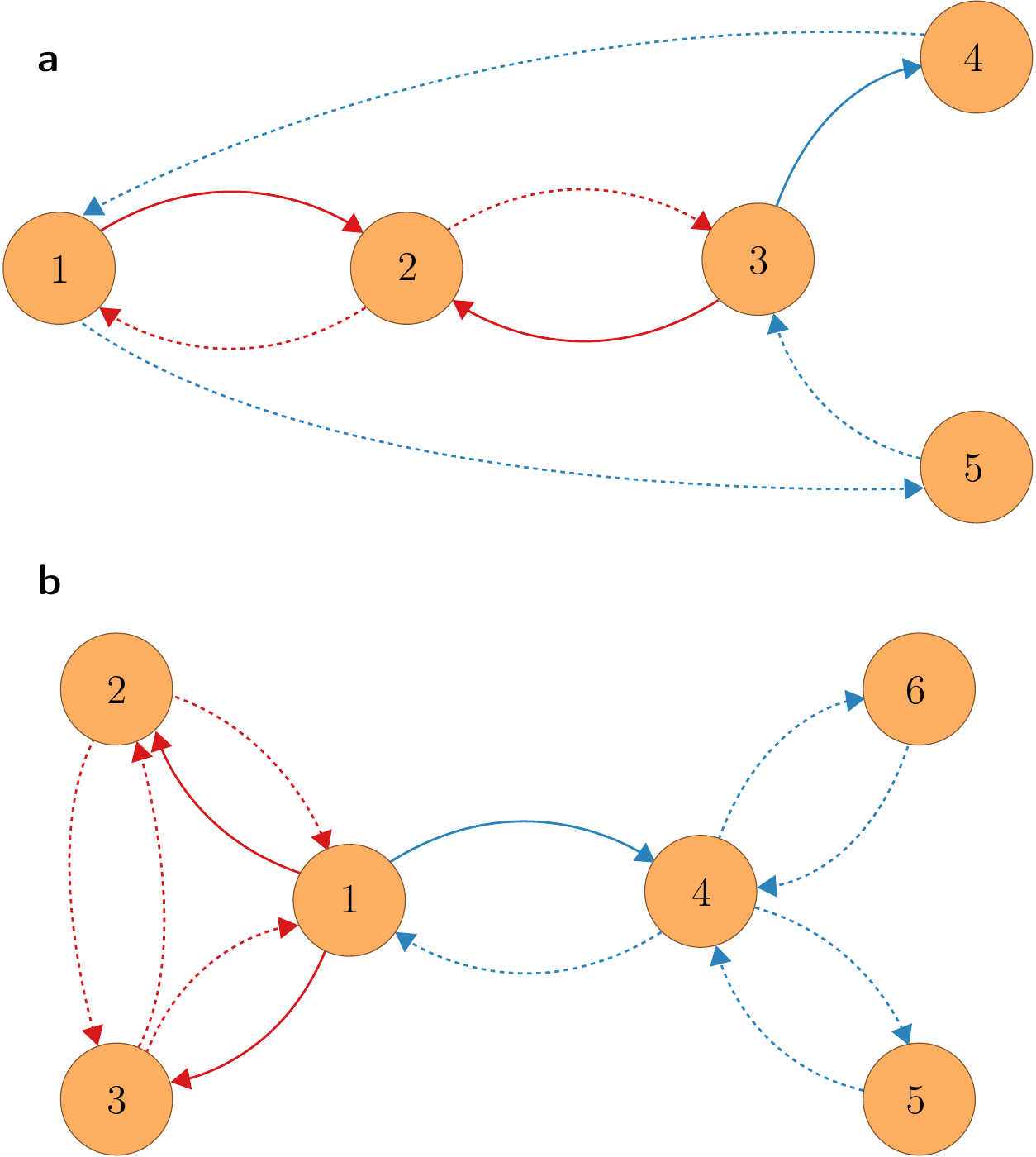}\\
\caption{\textbf{Schematic networks without triangles and with reciprocated links.} In \textbf{a}, there is no triangle and $r_3$ does not play a role. In \textbf{b}, all links are reciprocated and $r_2$ does not play a role. The stationary probability  $\pi$ increases/decreases with $r_2$ (\textbf{a}) or $r_3$ (\textbf{b}) for red and blue links respectively. For solid lines, this increase/decrease is predicted by the first-order perturbation in Supplementary Equation (\ref{perturbation}). For dashed lines, the probability of finding a walker on a memory node is uniform in the first-order approximation, and higher-order contributions are required to predict increase or decrease. \label{PRfig1}}
\end{figure}

We now turn to evaluating the effects of $r_2$, $r_3$, and $r_{3<}$ on PageRank. To do so, we use a perturbation analysis of  $\alpha$ close to $0$, where a local approximation of PageRank is valid. For the sake of simplicity, we consider the case of unweighted networks. We further assume that the graph is Eulerian, so that the known solution $\pi(\vv{jk})=1/L$  for $r_2 + r_3 + r_{3<}=1/3$ can be used as a baseline. In this case, it is straightforward to show that the dominant contribution to the stationary solution is
\begin{eqnarray}
\label{perturbation}
\pi(\vv{jk}) = \frac{1-\alpha}{L} + \frac{\alpha}{L} \sum_{i} A_{ij} \hat{p}(\vv{ij} \rightarrow \vv{jk}) + o(\alpha^2)\cr
 = \frac{1-\alpha}{L} + \frac{\alpha}{L} \sum_{i} A_{ij} \frac{A_{jk}  w(\vv{ij} \rightarrow \vv{jk})}{\sum_l A_{jl}  w(\vv{ij} \rightarrow \vv{jl})} + o(\alpha^2).
\end{eqnarray}
Intuitively, the PageRank of memory node $\vv{jk}$ increases when the bias of the random walk process favours transitions $\vv{ij} \rightarrow \vv{jk}$ over other transitions leaving $\vv{ij}$. Different types of scenarios are possible. As an illustration, we first focus on the role of $r_2$ for a network without triangles, such as the one of Supplementary Fig.\hs\ref{PRfig1}a. As expected, higher values of $r_2$ tend to favour reciprocated links. Higher-order contributions, corresponding to paths of length longer than 1, are expected to favour reciprocated links connected to many other reciprocated links, etc., due to the iterative nature of PageRank.  For instance, for memory node $\vv{12}$, one finds 
\begin{eqnarray}
\label{perturbation1}
\pi(\vv{12})  &=& \frac{1-\alpha}{8} + \frac{\alpha}{8} (\frac{1}{2} + \frac{8/10}{8/10 + 1/10}) + o(\alpha^2)\cr
&=& \frac{1}{8} (1 + \frac{7 \alpha}{18} ) + o(\alpha^2),
\end{eqnarray}
as shown in Supplementary Fig.\hs\ref{PRfig2}.
For $\vv{21}$, in contrast, the linear approximation is uniform 
\begin{eqnarray}
\label{perturbation2}
\pi(\vv{21})  =  \frac{1}{8} + o(\alpha^2),
\end{eqnarray}
and higher-order contributions are necessary to show that the bias boosts the PageRank. From (11) in the main manuscript, we see that physical nodes are important if they have many central incoming links. In the first example of Supplementary Fig.\hs\ref{PRfig1}, node 2 is very central while node 4 is not. 
In Supplementary Fig.\hs\ref{PRfig1}b, we also illustrate the role of $r_2$ on centrality, and show that high values of $r_2$ tend to favour links belonging to many triangles, as expected.

\begin{figure}[ht]
\centering
\includegraphics[width=0.9\columnwidth]{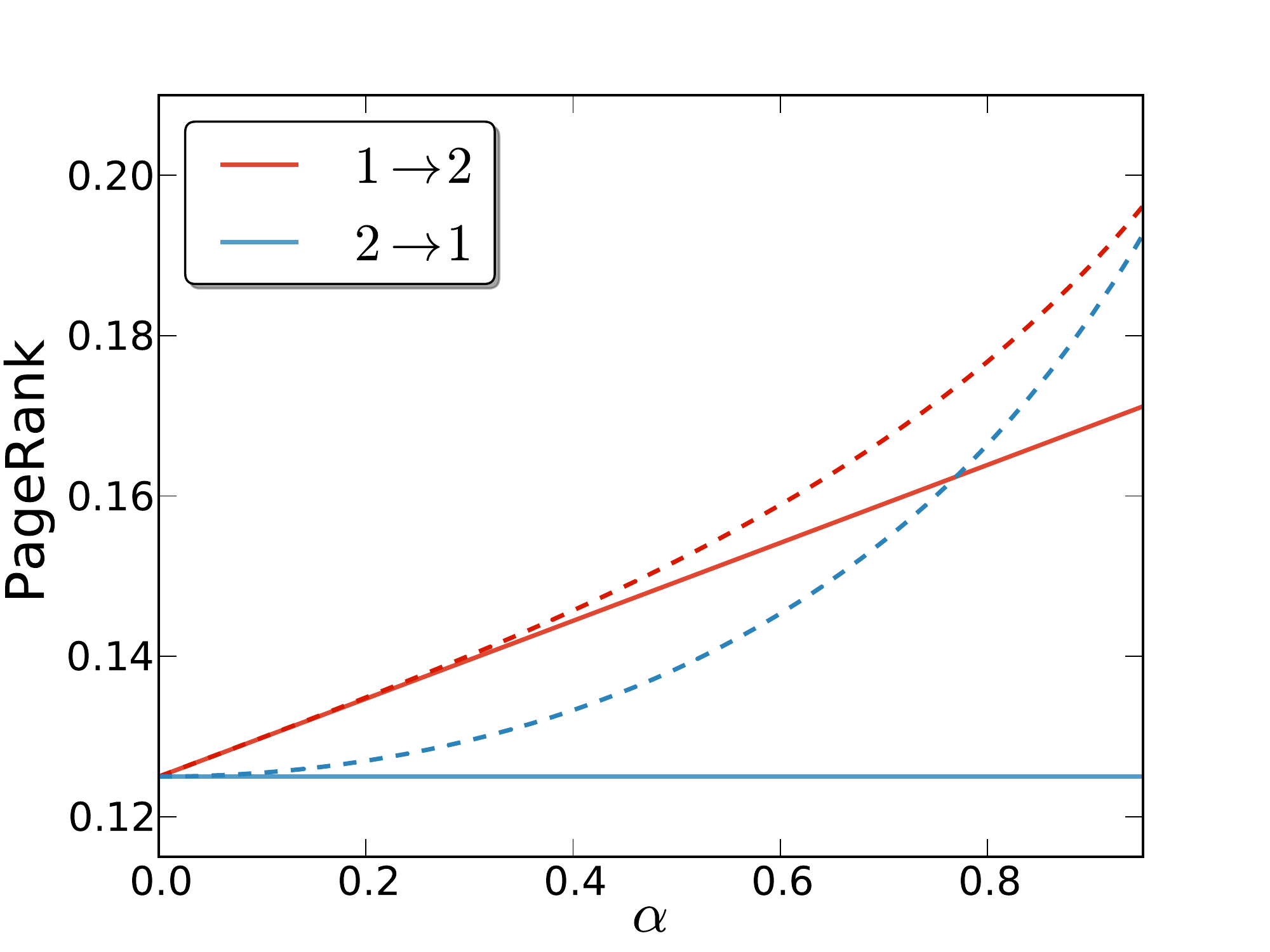}\\
\caption{\textbf{Linear approximation of memory effect on ranking.} PageRank as a function of $\alpha$ for edges going from $1$ to $2$ and from $2$ to $1$, illustrated in Supplementary Fig.\hs\ref{PRfig1}a, for $r_2=0.8$, $r_3=0.1$ and $r_{3<}=0.1$. Solid lines correspond to the linear approximation given by Supplementary Equation (\ref{perturbation1}) and Supplementary Equation (\ref{perturbation2}). \label{PRfig2}}
\end{figure}

\end{document}